\documentclass[pra,amsmath,amssymb,twocolumn,showpacs,superscriptaddress,longbibliography]{revtex4-1}
\usepackage{amsfonts,amsmath,amssymb,bm}
\usepackage{graphicx,graphics,float}
\usepackage{bm}
\usepackage{amssymb}
\usepackage{colordvi}
\usepackage{graphicx}
\usepackage{color}
\usepackage[colorlinks=true,linkcolor=blue,citecolor=blue,urlcolor=blue]{hyperref}
\usepackage{hyperref}
\usepackage{comment}
\usepackage{harpoon}
\usepackage{braket}
\usepackage{booktabs}
\usepackage{array}

\begin{document}
	
	\title{Non-Hermitian Effects in Dicke models}
	
	\author{Bin Jiang}
	\affiliation{Department of Modern Physics, School of Physical Sciences, University of Science and Technology of China, Hefei, 230026, China}
	\author{Yi-Yang Li}
	\affiliation{School of Physical Science and Technology \& Collaborative Innovation Center of Suzhou Nano Science and Technology, Soochow University, Suzhou 215006, China}
	\author{Junjie Liu}
	\affiliation{Department of Physics, International Center of Quantum and Molecular Structures, Shanghai University, Shanghai, 200444, China}
	\affiliation{Institute for Quantum Science and Technology, Shanghai University, Shanghai, 200444, China}
	\author{Chen Wang}
	\email{wangchen@zjnu.cn}
	\affiliation{Department of Physics, Zhejiang Normal University, Jinhua, Zhejiang, 321004, China}
	\author{Jian-Hua Jiang}
	\email{jhjiang3@ustc.edu.cn}
	\affiliation{School of Biomedical Engineering, Division of Life Sciences and Medicine, University of Science and Technology of China, Hefei 230026, China}
	\affiliation{Suzhou Institute for Advanced Research, University of Science and Technology of China, Suzhou, 215123, China}
	\affiliation{Department of Modern Physics, School of Physical Sciences, University of Science and Technology of China, Hefei, 230026, China}
	\affiliation{School of Physical Science and Technology \& Collaborative Innovation Center of Suzhou Nano Science and Technology, Soochow University, Suzhou 215006, China}
	
\date{\today}	

	\begin{abstract}
		
		The Dicke model, which describes the collective interaction between an ensemble of atoms and a single-mode photon field, serves as a fundamental framework for studying light-matter interactions and quantum electrodynamic phenomena. In this work, we investigate the manifestation of non-Hermitian effects in a generalized Dicke model, where two dissipative atom ensembles interact with a single-mode photon field. By applying the Holstein-Primakoff transformation, we explore the system in the semiclassical limit as a non-Hermitian Dicke model, revealing rich exceptional points (EPs) and diabolic points in such a system. We find that, by introducing the nonlinear saturation gain into an atomic ensemble, higher-order EP can be induced, leading to intriguing properties. Furthermore, if the system is extended to a one-dimensional chain, then the band topology will interplay with the non-Hermitian effect. In the quantum regime, we explore the quantum signature of EPs, noting that the conditions for their emergence are influenced by discrete photon numbers. We further study the transition from photon anti-bunching to bunching at a steady state, driven by non-Hermitian dynamics. Our findings deepen the understanding of non-Hermitian physics in light-matter interaction which is instructive for the design of advanced photonic and quantum systems.
		
	\end{abstract}
	
	\date{\today}
	
	\maketitle
	
	\section{Introduction}
	
	In the past decade, the progress in non-Hermitian physics has significantly advanced the understanding of nonequilibrium systems and impacted the development in other fields including photonics, metamaterials, quantum optics, and topological physics~\cite{gcma2022nrp,zhu2023,front_lee,RMP_Emil,rev_Ueda,rev_Saito}.
	Non-Hermitian systems are described by non-Hermitian Hamiltonians,
	of which the eigenvalues are real under the parity-time (\rm PT) symmetry, whereas complex eigenvalues occur once such symmetry is broken.
	An exceptional point (\rm EP) emerges at the symmetry-breaking transition, which is revealed as a spectral singularity where the eigenvalues coincide and the eigenvectors coalesce.
	
	Non-Hermitian light-matter interacting systems have attracted much attention both in semiclassical~\cite{cqwang2023aop,chlee2024apl,reg2019cp} and quantum regimes~\cite{xiaomin2016prl,pzoller2017nature,tjk2017nc,mueda2020aip,yhu2023prx}, where the light-matter interaction plays a crucial role in {\rm PT}-symmetry breaking. Such non-Hermitian hybrid systems yield versatile phenomena and applications, including the PT-symmetry controlled lasers~\cite{yu2021cp,oraz2022sa,hbwu2022pnas},
	non-Hermitian skin effects~\cite{boyan2022prl,yfchen2022aipx,jhjiang2024nc},
	non-Hermitian sensors~\cite{jcb2020prl,jf2023nature,Wiersig2020pr},
	and non-Hermitian phase transitions~\cite{pbl2019prl}.
	Meanwhile, the accompanied EP spectral singularity results in extraordinary topological features~\cite{ypwang2024prl,ac2024nc,gcma2022nrp}
	and unconventional quantum thermodynamics~\cite{mfeng2022nc,mfeng2023prl}.
	Non-Hermitian effects in the quantum regime as described by the Liouvillian theoretical framework have also been discussed~\cite{fnori2019pra,cc2018pra,dzueco2018pra}, where the Liouvillian {\rm EP} is generally distinct from its Hamiltonian counterpart.

	Dicke model~\cite{dicke1954pr}, a prototype model of light-matter interacting systems, characterizes the collective interaction between an ensemble of atoms and a single-mode photon field and is known to yield the superradiant phase transitions~\cite{pkirton2019aqt} and criticality-enhanced sensing~\cite{jmcai2021prl,mplenio2022prx}. A number of interesting theoretical works have been carried out to explore the influences of fertile light-matter interactions on the superradiant phase transition,
	including the counter-rotating-wave contributions~\cite{lambert2004prl}, the anisotropic couplings~\cite{das2023pra}, and the two-photon processes~\cite{garbe2017pra}. When non-Hermitian effects in the Dicke model are included via, e.g., nonreciprocal couplings~\cite{kpg2009pre,mb2023prl} or complex potential energy~\cite{zmli2023pra}, unconventional superradiant phase transition features have been revealed.
	
	Higher-order EPs, i.e., the coalescence of multiple eigenvalues (and the corresponding eigenvectors) at fine-tuned parameters, have recently attracted increasing research interest~\cite{HOEPtheory1,HOEP7,HOEP9}. So far, most studies mainly focus on second-order EPs in two-level systems as the simplest prototypes of EPs that can enhance sensing. With higher-order EPs, a system can be very sensitive to external perturbations and exhibit versatile quantum geometry, as found in various systems recently~\cite{HOEP7}. Generally, to achieve an $n$-th order EP, one needs $2n-2$ tunable physical parameters~\cite{HOEPtheory1}, which makes the realization of higher-order EPs rather challenging.
	
	In this work, we study the non-Hermitian effects in a generalized Dicke model where two individually dissipated atomic ensembles simultaneously interact with a single-mode photon field. In the semiclassical limit, i.e., the number of atoms (and photons) become extremely large, such a generalized Dicke model can be approximately described by a non-Hermitian Dicke Hamiltonian which can be obtained via the Holstein-Primakoff transformation. We find that this non-Hermitian Dicke Hamiltonian have rich non-Hermitian phases. We further extend the non-Hermitian Dicke model by considering the nonlinear saturation gain in one of the atomic ensembles which yields even higher-order EPs without increasing the degrees of freedom. By extending the non-Hermitian Dicke system to a one-dimensional (1D) Dicke chain, we find the intriguing interplay between non-Hermiticity and topology. In the quantum limit, i.e., when the number of photons is small, the quantum signature of EPs is discussed, featuring the photon-number-dependent EPs. Furthermore, the influence of non-Hermiticity on the photon statistics, e.g., photon bounching and anti-bunching, is analyzed. Finally, we discuss possible physical systems that realize the non-Hermitian Dicke model.
	
	The remaining of the paper is organized as follows:
	Section~\ref{ClassicalDickeModel} contains the result for the semiclassical non-Hermitian Dicke model, analyzing the high-order EPs, nonlinear gain, dynamics, and topological phase transformation, accordingly.
	Section~\ref{QuantumDickeModel} explores the quantum counterpart of non-Hermitian Dicke model.
	Section~\ref{PossibleRealizations} conceives possible realizations of non-Hermitian Dicke model both in semiclassical and quantum regimes.
	Section~\ref{Summary} gives the conclusion and discussions for this work.

	\section{Non-Hermitian Dicke Model in the semiclassical regime} \label{ClassicalDickeModel}
	We now introduce the ``semiclassical'' version of the non-Hermitian Dicke model and subsequently analyze the high-order EPs.

	\begin{figure}[htb]
		\begin{center}
			\centering
			\includegraphics[width=8.5cm]{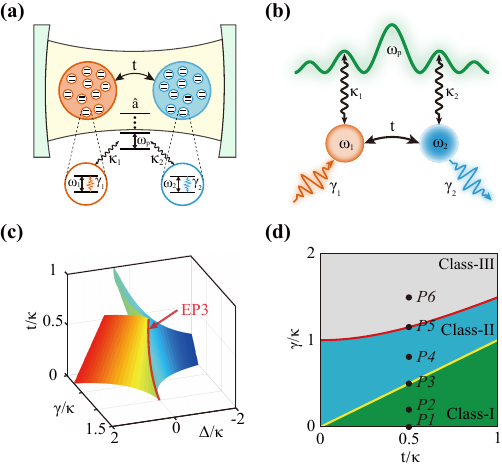}
			\caption{
				(a) Schematic of the non-Hermitian quantum Dicke model, where two coupled atom ensembles interact with a quantized photon field.
				The collective angular momentum operator consisted of $N_i$ spin-1/2 atoms with the transition energy $\omega_i~(i=1,2)$,
				while $\hat{a}$ denotes the annihilation operator for a photon with the frequency $\omega_p$.
				$\kappa_i$ denotes the coupling strength between each atom and the photon field,
				and $t$ indicates the collective angular momentum coupling strength.
				Positive(negative) $\gamma_i$ denotes the gain(loss) strength of the $i$th collective angular momentum.
				(b) Schematic of the non-Hermitian Dicke model in the semiclassical regime, where two coupled atom (depicted as orange and blue solid circles) ensembles interact with a single-mode photonic field (green wavy line).
				$t$ represents the coupled strength between the atomic ensembles,
				while $\kappa_{i}$ represents the coupling strength between each atom and the photonic field,
				and $\gamma_{i}$ represents the gain (or loss) rate of the $i$th atom for $i=1,2$.
				(c) EP2 and EP3 in 3D parameter spaces $\left \{ \Delta/\kappa,\gamma/\kappa,t/\kappa \right \}$.
				A colorful map represents the surface of EP2,
				and the intersection of two EP2 surface branches (half-transparent red line) outlines the trajectory of EP3.
				(d) The EP phase diagram of the semiclassic Dicke model in 2D parameter spaces $\left \{ \gamma/\kappa, t/\kappa \right \}$.
				The green, blue, and gray regions represent classes I, II, and III EPFPs, respectively.
				The solid red curve marks the formation of EP3,
				while the solid yellow line indicates the transition of EP2.
				The six black dots, labeled from \emph{P1} to \emph{P6}, along a vertical line,
				represent specific EPFPs, as illustrated in Fig.~\ref{fig2}.
			}
			\label{fig1}
		\end{center}
	\end{figure}

	\subsection{Model}
	The quantum Dicke model, which is composed of two coupled atom ensembles interacting with a quantized photon field, reads as
	\begin{eqnarray}
		\hat{H}_{\rm Dicke}&=&\sum_{j=1,2}\omega_j\hat{J}^z_j
		+\omega_p\hat{a}^\dag\hat{a}
		+\frac{t}{\sqrt{N_1N_2}}(\hat{J}^+_1\hat{J}^-_2+\hat{J}^-_1\hat{J}^+_2)\nonumber\\
		&&+\sum_{j=1,2}\frac{\kappa_j}{\sqrt{2N_j}}(\hat{a}^\dag\hat{J}^-_j+h.c.),
	\end{eqnarray}
	where $\hat{J}^\alpha_i=\sum^{N_i}_{n=1}\hat{s}^n_\alpha~(\alpha=z,+,-)$ describe the collective angular momentum operator consisting of $N_i$ spin-1/2 atoms, with the transition energy $\omega_i$,
	$\hat{a}^\dag(\hat{a})$ creates(annihilates) one photon with the frequency $\omega_p$,
	$\kappa_i$ denotes the coupling strength between atomic ensembles and the photon field,
	and $t$ shows the collective angular momentum coupling strength.
	
	We consider the influence of non-Hermitian effect on the quantum Dicke model via separate gain and loss processes on atom ensembles, the Hamiltonian is described as
	\begin{eqnarray}
		\hat{H}_{\rm nDicke}=\hat{H}_{\rm Dicke}+\sum_{j=1,2}\frac{i\gamma_j}{N_j}\hat{J}^+_j\hat{J}^-_j,
	\end{eqnarray}
	where positive(negative) $\gamma_i~(i=1,2)$ denotes the gain(loss) strength of the $i$th collective angular momentum to break the hermiticity of the Dicke model.
	The quantum dissipation of the photon field is assumed to be comparatively ignored, which can be restored straightforwardly.
	
	Then we include the Holstein-Primakoff transformation~\cite{HPtransformation} to the collective angular momentum operators, \emph{i.e.}
	$\hat{J}^+_i=\hat{b}^+_i\sqrt{N_i-\hat{b}^+_i\hat{b}_i}$,
	$\hat{J}^-_i=\sqrt{N_i-\hat{b}^+_i\hat{b}_i}\hat{b}_i$,
	and
	$\hat{J}^z_i=\hat{b}^+_i\hat{b}_i-N_i$.
	Considering weak excitation of collective angular momentum operators ${\langle}\hat{b}^+_i\hat{b}_i{\rangle}{\ll}N_i$ with huge numbers of atoms,
	the angular momentum operators can be approximately expressed as
	$\hat{J}^+_i=\sqrt{N_i}\hat{b}^+_i$,
	$\hat{J}^-_i=\sqrt{N_i}\hat{b}_i$,
	and
	$\hat{J}^z_i=\hat{b}^+_i\hat{b}_i-N_i$.
	Thus the  Hamiltonian of the Dicke model in the semiclassical version of non-Hermitian Dicke model as shown in Fig.~\ref{fig1}(b), can be reexpressed as
	\begin{eqnarray}~\label{cdicke1}
		\hat{H}_{\rm cDicke}&=&
		\sum_{j=1,2}(\omega_j+i\gamma_j)\hat{b}^\dag_j\hat{b}_j+\omega_0\hat{a}^\dag\hat{a}\\
		&&+t(\hat{b}^\dag_1\hat{b}_2+\hat{b}_1\hat{b}^\dag_2)+
		\sum_{j=1,2}\frac{\kappa_j}{\sqrt{2}}(\hat{b}^\dag_j\hat{a}+h.c.)\nonumber.
	\end{eqnarray}

	\subsection{High-order EPs}
	
	In this section, we investigate high-order EPs where multiple eigenstates coalesce in a three-mode non-Hermitian system within the semiclassical Dicke model.
	The emergence of multiple EPs and their interactions are summarized in a phase diagram. The formation of EP3 (the coalescence of two EP2) and the transition of EP2, two curves in a parameter space dividing the phase space into three regions, each with a unique EP formation pattern (EPFP)~\cite{HOEP6}.
	Furthermore, we assess the order of singularities of EPs by employing the concept of phase rigidity, confirming EP3(EP2) as a third(second)-order singularity.
	
	High-order EPs~\cite{HOEP1, HOEP2}, i.e.
	non-Hermitian degeneracies,
	denote particular points in parameter spaces where more than two eigenvalues and the corresponding eigenvectors simultaneously coalesce.
	Such high-order EPs have already been theoretically~\cite{HOEPtheory1, HOEPtheory2} and experimentally explored in photonic crystals~\cite{HOEPphoton1, HOEPphoton2, HOEPphoton3, HOEPphoton4}, phononic crystals~\cite{HOEP2, HOEP6} and  hybrid quantum systems~\cite{HOEPquantum1, HOEPquantum2}, exhibiting enhanced detection sensitivity~\cite{HOEP7, HOEP8, HOEP9} and spontaneous emission~\cite{HOEPemission1, HOEPemission2}.
	
	Here we specifically consider two identical atom ensembles, \emph{i.e.} $\omega_{1}=\omega_{2}=\omega_{a}$ and $\kappa_{1}=\kappa_{2}=\kappa$.
	Under such condition
	we directly confirm that there is no third-order EP (EP3) if introducing only loss (or gain) in our semi-classical Dicke model at Eq.~(\ref{cdicke1}) (please see in the Supplementary materials).
	To recover higher-order EPs in our model, we should separately include balanced gain and loss processes at two atom ensembles, \emph{e.g.}, $\gamma_{1}=-\gamma_{2}=\gamma$.
	It is known that the global frequency shift is irrelevant to the eigensolution of $\hat{H}_{\rm cDicke}$.
	Thus we reexpress the semiclassical Dicke model as
	$\hat{H}_{\rm cDicke}=[\hat{b}^\dag_1,\hat{a},\hat{b}^\dag_2](\omega_{p}\hat{I}_{3\times3}+\mathcal{H})[\hat{b}_1,\hat{a},\hat{b}_2]^T$,
	where the corresponding Dicke matrix is specified as
	\begin{eqnarray}
		\mathcal{H}=
		\begin{bmatrix}
			\Delta + i\gamma & \kappa/\sqrt{2} & t\\
			\kappa/\sqrt{2}  & 0 &\kappa/\sqrt{2} \\
			t & \kappa/\sqrt{2} & \Delta - i\gamma
		\end{bmatrix}, \label{Eq4}
	\end{eqnarray}
	with $\Delta=\omega_{a}-\omega_{p}$  the frequency detuning between the atom ensemble and the photon field. {In the following analysis, $\kappa$ is set as the energy unit.}

	\begin{figure}[htb]
		\begin{center}
			\centering
			\includegraphics[width=8.5cm]{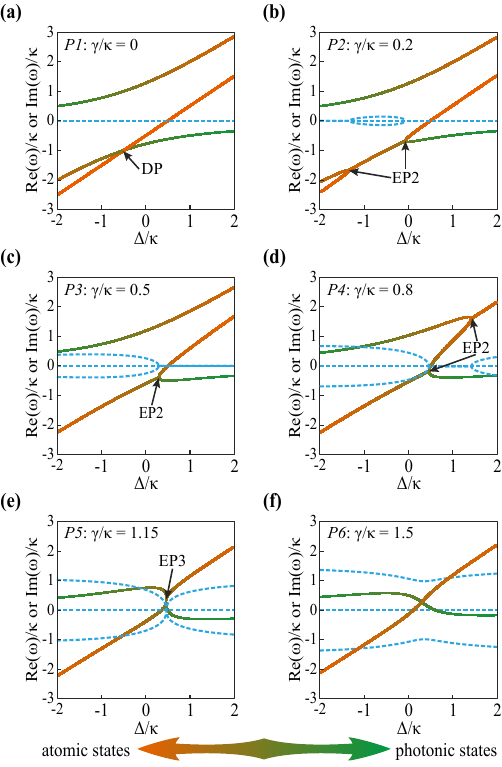}
			\caption{ Eigenvalues as a function of frequency detuning $\Delta/\kappa$ from points \emph{P1} to \emph{P6} within the phase diagram in Fig.~\ref{fig1}(d), accordingly. The solid lines represent the real part of the eigenvalues, while the dashed lines represent the imaginary part. The real part of the eigenvalues is colored according to the content of the eigenstates, as labeled by the colorscheme at the bottom of the figure: The red color stands for the atomic states, the green color denotes the photonic states, while a mixed state is represented by the color in between. Thus, the color change represents the change in the content of the eigenstates. All parameters are normalized by $\kappa$ and $t/\kappa=0.5$ is adopted for all panels.
				The balanced gain (loss) strength $\gamma/\kappa$ is given in the top left corner of each panel. Each exceptional point is labeled to show its type according to the main text.}
			\label{fig2}
		\end{center}
	\end{figure}

	We include Eq.~(\ref{Eq4}) to investigate EPs in Fig.~\ref{fig1} by tuning $t/\kappa$, $\gamma/\kappa$ and $\Delta/\kappa$, where fertile EPs are unraveled.
	First, we analyze the emergence of second-order EP (EP2),
	indicating that the eigenvalues spectra in Eq.~(\ref{Eq4}) include two conjunctive eigenvalues $\omega_{EP2}$ (with their algebraic multiplicity to be 2 and geometric multiplicity to be 1) and another single eigenvalue $\omega_{3}$,
	which can be numerically obtained.
	Utilizing the Vieta's formulas for cubic equations, we obtain the parameter equations of $\Delta$ and $\gamma$ regarding $\omega_{EP2}$ (the details of derivations please see in Appendix A)
	\begin{subequations}
		\begin{eqnarray}
			\Delta=\frac{t\kappa^2+2\omega_{EP2}^3}{\kappa^2+2\omega_{EP2}^2}, \\
			\gamma=\frac{\sqrt{\kappa^2+\omega_{EP2}^2}(\kappa^2+2t\omega_{EP2})}{\kappa^2+2\omega_{EP2}^2}, \\
			\omega_{3}=\frac{2(t-\omega_{EP2})\kappa^2}{\kappa^2+2\omega_{EP2}^2}. \label{Eq38c}
		\end{eqnarray} \label{EP2}
	\end{subequations}
	Figure~\ref{fig1}(c) presents the EP2 surface in parameter spaces obtained from above Eq.~(\ref{EP2}).
	The same eigenvalues are dying identical colors consisting of iso-frequency lines. The intersection of two EP2 surface branches (semitransparent red line in Fig.~\ref{fig1}(c)) with the same eigenvalues at each point, merges two EP2 into an astonishing EP3.
	The emergence of EP3 also can be understood through the crossing between EP2 and another single eigenvalue $\omega_{3}$.
	Under such a search bound $\omega_{EP2}=\omega_{3}=\omega_{EP3}$, the Eq.~(\ref{EP2}) can be reduced to (the details of derivations please see in Appendix A)
	\begin{subequations}
		\begin{eqnarray}
			\Delta = \frac{3}{2}\omega_{EP3},\\
			\gamma = \frac{1}{\kappa^2}(\omega_{EP3}^2+\kappa^2)^\frac{3}{2}, \\
			t=\frac{3}{2}\omega_{EP3}+\frac{\omega_{EP3}^3}{\kappa^2}.
		\end{eqnarray} \label{EP3}
	\end{subequations}
	Therefore, the signal of higher-order EPs can be
	explicitly observed by tuning the system parameters.
	
	Then we study eigenvalues in Eq.~(\ref{Eq4}) as a function of $\Delta/\kappa$ and analyze the distribution of EPs by tuning $t/\kappa$ and $\gamma/\kappa$.
	Fig.~\ref{fig1}(c) shows that depending on the parameters $t/\kappa$ and $\gamma/\kappa$, different combinations of EP2 and EP3 consist of different EPFPs.
	And figure ~\ref{fig1}(d) shows a birdview of EP phase diagram.
	Specifically, there are three regions marked as class-I, II, and III, with their boundaries highlighted by a solid yellow line and a solid red line.
	Interestingly, these EPs classify the regime into three regions, each owning a unique EPFP.
	
	To present the EPFP in each region, the inter-site coupling is given a special value of $t/\kappa=0.5$ and then we increase $\gamma/\kappa$ gradually from 0 (Point \emph{P1}) to 1.5 (Point \emph{P6}), as marked by black points in Fig.~\ref{fig1}(d).
	The point \emph{P1} ($\gamma/\kappa=0$) turning off the balance loss and gain is a Hermitian case, whose spectrum is shown in Fig.~\ref{fig2}(a).
	There is a linear crossing diabolic point (DP) at $\Delta/\kappa = t/\kappa$.
	We stress that such a linear crossing DP in Hermitian system is essentially different from EPs in non-Hermitian systems, where the singularity(the coalescence of multiple eigenstates simultaneously) is found.
	
	Once we slightly break hermiticity in our Dicke model by introducing a finite balance gain and loss, \emph{e.g.} $\gamma/\kappa=0.2$, the linear crossing DP in Hermitian system will split into two EP2s with opposite chirality, which gives a typical EPFP of class-I (Point \emph{P2}), as shown in Fig.\ref{fig2}(b).
	Here, we define the chirality of EPs according to their eigenstates.
	At an EP, two (or more) eigenvalues and their corresponding eigenvectors coalesce, leaving only one independent eigenstate.
	The behavior of this eigenstate under small perturbations,
	including how it rotates or evolves around the EP in parameter space,
	determines the chirality of the EP.
	This chirality can indeed be viewed similarly to the polarization of light, with the eigenvector rotating either clockwise (right-handed chirality) or counterclockwise (left-handed chirality) as parameters (such as system Hamiltonian variables) are varied.
	The direction of rotation indicates the nature of the degeneracy at the EP. Importantly, since only one eigenvector remains at the EP, its response to perturbations encapsulates the geometric phase and topological features of the EP, linking chirality with the topological properties of the system.
	
	In the turning process of increasing $\gamma/\kappa$ continuously, it is found that the occurring positions of two EP2 are in opposite directions, and the left-one part disappears (or moves to negative infinite far way from another EP2) when $\gamma/\kappa = t/\kappa = 0.5$, as shown in Fig.~\ref{fig2}(c).
	The value of $\omega_{EP2}$ can be greater than $\omega_{3}$ only if $t/\gamma$ is less than 1, which can be seen from Eq.~(\ref{EP2}).
	Therefore, the yellow curve $t/\gamma = 1$ in Fig.~\ref{fig1}(d) is a critical middle phase during the process of increasing $\gamma/t$.
	The yellow line $\gamma = t$ in Fig.~\ref{fig1}(d) separates class-I from class-II, marking the transition of an EP2 from the lower energy gap to the higher one.
	This phase boundary is of particular importance because it not only shifts the location of the EP2 but also flips its chirality.
	Below the yellow line in the region of class-I, two EP2 appear in the same energy gap with opposite chirality while above the yellow line in the region of class-II, two EP2 with the same chirality in adjacent energy gaps and the corresponding  EPFP are shown in Fig.~\ref{fig2}(d).
	Upon a further increase in $\gamma/\kappa$, the system will meet the red line in Fig.~\ref{fig1}(d). As shown in Fig.~\ref{fig2}(e), the configuration of Point \emph{P5} on the red line always exists a EP3 at one particular value of $\Delta/\kappa$.
	The analytical parameter expressions of $\Delta$, $\gamma$ and $t$ regrading the eigenvalue at EP3 $\omega_{EP3}$ are given in Eq.~(\ref{EP3}).
	
	Interestingly, we also notice that the red curve approaches asymptotically the yellow red line, \emph{i.e.} ${\lim_{t\to +\infty}\gamma/t=1}$, which can be seen from Eq.~(\ref{EP3}).
	Eventually, the increasing of $\gamma/\kappa$ brings the Dicke model into region class-III, where the system will thoroughly lose PT symmetry and the real part of the eigenvalue of EP3 still keeps degenerating, while the imaginary parts are expected to be splitting (Point \emph{P6} in Fig.~\ref{fig1}(d) and its corresponding eigenvalues spectra in Fig.~\ref{fig2}(f)).
	
	Before delving into the singularities of EPs,
	it's important to summarize the ``chemical'' interactions that lead to the coalescence of two or more EPs.
	The emergence of an EP3 can be understood as the merger of two EP2s in adjacent energy gaps with the same chirality.
	In contrast, when EP2s at the same energy gap with the same chirality collide,
	they merge into a linear crossing EP2,
	a phenomenon noted in Ref.~\cite{HOEP6}.
	Conversely, a linear crossing DP in Hermitian systems forms through the merger of two EP2s within the same energy gap but with opposite chirality,
	which can be observed during the reversal detuning process from \emph{P2} to \emph{P1}.
	It is important to note that the linear crossing EP2 in non-Hermitian systems is distinct from the linear crossing DP in Hermitian systems,
	as the latter does not exhibit any singularity.

	\begin{figure}[htb]
		\begin{center}
			\centering
			\includegraphics[width=8.5cm]{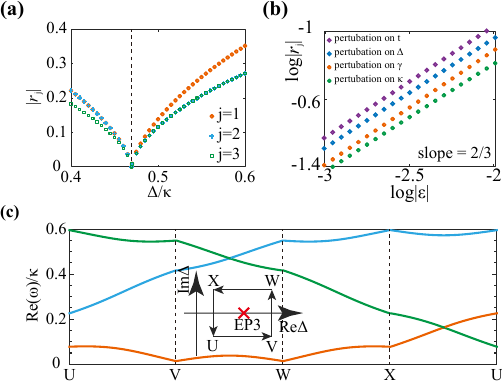}
			\caption{
				(a) The absolute value of phase rigidity of all the eigenvectors as functions of frequency detuning $\Delta$ for the Point \emph{P5} in the Phase diagram shown in Fig.~\ref{fig2}(e).
				Here $j = 1,2$ or $3$ labeled energy levels ranked according to the real part of eigenvalues.
				(b) Phase rigidity versus perturbations near the EP3 in Fig.~\ref{fig2}(e).
				The slope $s=2/3$ in the double logarithmic axis indicates a second-order exceptional point.
				External perturbations on $t$, $\Delta$, $\gamma$, or $\kappa$ are plotted in purple, blue, orange, and green circles, respectively.
				(c) Eigenfrequency trajectories for looping around the EP3 in Fig.~\ref{fig2}(e) in the counterclockwise direction ($U \rightarrow V\rightarrow W\rightarrow X \rightarrow U$) as shown in the inset.
				The inset shows the looping path in the complex-$\Delta$ plane, in which EP3 is located inside the loop.
			}
			\label{aroundEP3}
		\end{center}
	\end{figure}

	\begin{figure}[htb]
		\begin{center}
			\centering
			\includegraphics[width=8.5cm]{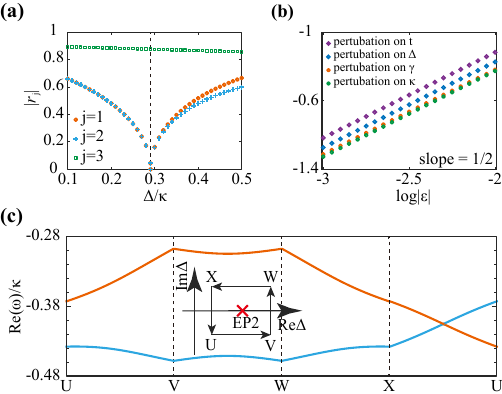}
			\caption{
				(a) The absolute value of phase rigidity of all the eigenvectors as functions of frequency detuning $\Delta$ for the Point \emph{P3} in the Phase diagram shown in Fig.~\ref{fig2}(c). Here $j = 1,2$ or $3$ labeled energy levels ranked according to the real part of eigenvalues.
				(b) Phase rigidity versus perturbations near the EP2 in Fig.~\ref{fig2}(c). The slope $s=1/2$ in the double logarithmic axis indicates a second-order exceptional point. External perturbations on $t$, $\Delta$, $\gamma$ or $\kappa$ are plotted in purple, blue, orange, and green circles, respectively.
				(c) Eigenfrequency trajectories for looping around the EP2 in Fig.~\ref{fig2}(c) in the counterclockwise direction ($U \rightarrow V\rightarrow W\rightarrow X \rightarrow U$) as shown in the inset. The inset shows the looping path in the complex-$\Delta$ plane, in which EP2 is located inside the loop.
			}
			\label{aroundEP2}
		\end{center}
	\end{figure}

	To identity the singularity order in the red line of the phase graph in Fig.~\ref{fig1}(d), we plot the absolute value of phase rigidity~\cite{PhaseRidity} in Fig.~\ref{aroundEP3}(a), defined as
	\begin{eqnarray}
		r_{j}=\frac{\langle\psi_{j}^{L}|\psi_{j}^{R}\rangle}{\sqrt{\langle\psi_{j}^{L}|\psi_{j}^{L}\rangle}\sqrt{\langle\psi_{j}^{R}|\psi_{j}^{R}\rangle}},
	\end{eqnarray}
	for each state $j$ as a function of $\Delta/\kappa$ for point \emph{P5} in the phase diagram.
	Phase rigidity is a quantity to capture the nonequivalence between right and left eigenstates in non-Hermitian system.
	It is clear that $|r_{j}|$ vanishes for all states at the EP3 at $\Delta/\kappa\approx0.47$, indicating a coalescence of three defective states.
	
	To investigate the critical behaviors of EP3 in Fig.~\ref{fig2}(e), we include an external perturbation factor $\varepsilon$ onto various parameters. The response of phase rigidity $|r_{j}|$ near EP3 is proportional to $\epsilon^{\frac{2}{3}}$ as shown in Fig.~\ref{aroundEP3}(b), which is independent on perturbation on concrete parameters. To investigate the physical significance of the slope $s=2/3$ of phase rigidity near an EP3,
	we perform an adiabatic process encircling the EP3 in the complex-$\Delta$ plane in a counterclockwise direction ($U \rightarrow V\rightarrow W\rightarrow X \rightarrow U$),
	as illustrated in the inset of Fig.~\ref{aroundEP3}(c).
	The trajectories of the real parts of the three eigenfrequencies along the path are shown in Fig.~\ref{aroundEP3}(c),
	from which we can see that the state $3$ is swapped with the other two states successively and reaches state $1$ after two swaps.
	Upon completing one full encirclement, the eigenvalue permutations can be represented by the permutation operation $P_{3}$:~$(123)\rightarrow(312)$.
	Notably, it takes three such cycles to return all eigenvalues to their original positions, as $P_{3}^3=\mathbb{1}$  denotes that the identity permutation is achieved: $\mathbb{1}$:~$(123)\rightarrow(123)$.
	The calculation of geometric phase by using parallel transport method~\cite{ParellelTransport} gives a geometric phase of $\pm2\pi$ after three cycles,
	consistence with the slope of $s=2/3$ found in Fig.~\ref{aroundEP3}(b).
	We hence directly confirm that the red line of the phase graph in Fig.~\ref{fig1}(d) is indeed a high-order EP3 singularity line.
	
	To further identify the singularity order in the yellow line of the phase graph in Fig.~\ref{fig1}(d),
	the absolute value of phase rigidity for each state $j$ as a function of $\Delta/\gamma$ for point \emph{P3} is shown in Fig.~\ref{aroundEP2}(a).
	It is clear that $|r_{j}|$ vanishes for states $j=1$ and $2$ at the EP2 at $\Delta/\kappa\approx0.29$, indicating a coalescence of two defective states.
	To investigate the critical behaviors of EP2 in Fig.~\ref{fig2}(c), we include an external perturbation factor $\varepsilon$ onto various parameters. The response of phase rigidity $|r_{j}|$ near EP2 is proportional to $\epsilon^{\frac{1}{2}}$ as shown in Fig.~\ref{aroundEP2}(b), which is independent on perturbation on concrete parameters. To investigate the physical significance of the slope $s=1/2$ of phase rigidity near an EP2,
	we perform an adiabatic process encircling the EP2 in the complex-$\Delta$ plane in a counterclockwise direction ($U \rightarrow V\rightarrow W\rightarrow X \rightarrow U$),
	as illustrated in the inset of Fig.~\ref{aroundEP2}(c).
	The trajectories of the real parts of the two eigenfrequencies along the path are shown in Fig.~\ref{aroundEP2}(c),
	from which we can see that the state $1$ is swapped with state $2$.
	Upon completing one full encirclement, the eigenvalue permutations can be represented by the permutation operation $P_{2}$:~$(123)\rightarrow(213)$.
	Notably, it takes two such cycles to return all eigenvalues to their original positions, as $P_{2}^2=\mathbb{1}$  denotes that the identity permutation is achieved: $\mathbb{1}$:~$(123)\rightarrow(123)$.
	The calculation of geometric phase by using parallel transport method~\cite{ParellelTransport} gives a geometric phase of $\pm\pi$ after one cycle,
	consistence with the slope of $s=1/2$ found in Fig.~\ref{aroundEP2}(b).
	We hence directly confirm that the yellow line of the phase graph in Fig.~\ref{fig1}(d) is indeed an EP2 singularity line.

	\subsection{Nonlinear saturation gain}
	
	\begin{figure}[htb]
		\begin{center}
			\centering
			\includegraphics[width=8.5cm]{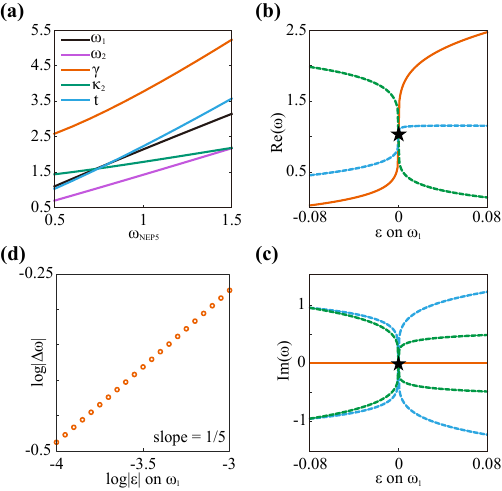}
			\caption{(a) The trajectory of steady solution $\omega_{s}$ of NEP5 with $\kappa_1=1$.
				(b) Real part of eigenvalues near NEP5 by adding perturbation $\epsilon$ onto $\omega_{1}$.
				Solid orange lines represent stable modes, and dashed lines are $4$ auxiliary modes.
				The dashed lines with the same color indicate the conjugate pair of auxiliary modes.
				The black pentacle indicates a NEP5.
				(c) Image part of eigenvalues near NEP5,
				where the stable mode has zero imaginary part,
				and four auxiliary modes appear with conjugate imaginary part.
				(d) The bilogarithmic plot of strain of eigenvalues $|\Delta\omega|$ by perturbing $\omega_1$.
				Here the circles come from the stable mode in panels (b) and (c), standing in a straight line with a slope of 1/5.
				In panels (b-d), the parameters we used for NEP5 can be obtained from Eq.~(\ref{SuppFiveOrderCharac}) with steady solution $\omega_{s}=1$.}
			\label{NEP5}
		\end{center}
	\end{figure}

	Now we consider the nonlinear gain case in Eq. (\ref{Eq4}). The linear gain above is replaced by nonlinear saturation gain in the first atom, which depends on the wave amplitude. The nonlinear stationary Schr\"odinger equation is expressed as
	\begin{eqnarray}~\label{nlcd1}
		\hat{H}_{\rm NL}(|\psi_{1}^{R}|)|\psi^{R}\rangle=\omega_s|\psi^{R}\rangle, \label{NLschrodinger}
	\end{eqnarray}
	where $\omega_s$ is the eigenfrequency,
	$|\psi^{R}\rangle\equiv(\psi_{1}^R, \psi_{p}^R, \psi_{2}^R)^T$ is the right eigenvector with superscript $T$ short for matrix transpose, and $\psi_{i}^R$($\psi_{p}^R$) is the wave amplitude of the $i$-th atom (photon).
	The nonlinear Hamiltonian is described as
	\begin{eqnarray}
		\hat{H}_{\rm NL}(|\psi_{1}^{R}|)=
		\begin{bmatrix}
			\omega_{1} + iG_{1}(|\psi_{1}^{R}|) & \kappa_{1} & t\\
			\kappa_{1}  & 0 &\kappa_{2} \\
			t & \kappa_{2} & \omega_{2} - i\gamma
		\end{bmatrix}. \label{Eq8}
	\end{eqnarray}
	In optical (e.g., laser) systems,
	the nonlinear saturation gain  can generally be specified as
	$G_{1}(|\psi_{1}^{R}|)={\alpha}/({1+|\psi_{1}^R|^2})-\beta$ \cite{nonlinear_gain1, nonlinear_gain2},
	where $\alpha$ represents the pumping strength and $\beta$ represents the intrinsic loss strength.
	
	To find steady eigenfrequencies,
	we solve the nonlinear Schr\"odinger Eq.~(\ref{NLschrodinger}) and obtain characteristic polynomials,
	\begin{eqnarray}
		\mathrm{Det}(\hat{H}_{NL}-\omega_{s}\hat{I}_{3\times3})=0.
		\label{SuppCharacPolyNL}
	\end{eqnarray}
	We split the real and imaginary parts of Eq.~(\ref{SuppCharacPolyNL}) in the left-hand side.
	The real part is
	\begin{eqnarray}
		\label{SuppNLRealPart111}
		\kappa_1^2(\omega_2-\omega_{s})+\kappa_2^2(\omega_1-\omega_{s})-2\kappa_{1}\kappa_{2}t+ \\ \nonumber
		\omega_{s}[(\omega_{1}-\omega_{s})(\omega_{2}-\omega_{s})+(G_{1}\gamma-t^2)]=0,
	\end{eqnarray}
	and the imaginary part is
	\begin{eqnarray}
		-\kappa_{1}^2\gamma+\kappa_{2}^{2}G_{1}+\omega_{s}[G_{1}(\omega_{2}-\omega_{s})-\gamma(\omega_1-\omega_{s})]=0. \label{SuppNLImagPart}
	\end{eqnarray}
	Combined with Eq.~(\ref{SuppNLRealPart111}) and Eq.~(\ref{SuppNLImagPart}),
	we eliminate the $G_{1}(|\psi_{1}^{R}|)$ and then obtain a five-order polynomial concerned about steady eigenfrequency $\omega_{s}$,
	\begin{eqnarray}
		\rho(\omega_{s})=-\omega_{s}^5+x_{4}\omega_{s}^4+x_{3}\omega_{s}^3+x_{2}\omega_{s}^2+x_{1}\omega_{s}+x_{0},
		\label{SuppFiveOrderCharac}
	\end{eqnarray}
	where the coefficients are $x_0=\kappa_{1}^2\kappa_{2}^2\omega_{2}-2\kappa_{1}\kappa_{2}^3t+\omega_1\kappa_{2}^4$,
	$x_1 = -\kappa_{1}^2\kappa_{2}^2+\kappa_{1}^2\gamma^2+\kappa_{1}^2\omega_{2}^2-2\kappa_{1}\kappa_{2}t\omega_2-\kappa_{2}^4-\kappa_2^{2}t^2+2\omega_1\kappa_{2}^2\omega_2$,
	$x_2 = -2\kappa_{1}^2\omega_{2}+2\kappa_{1}\kappa_{2}t-2\kappa_{2}^2\omega_{2}-2\omega_{1}\kappa_{2}^2+\omega_{1}\gamma^2-t^2\omega_{2}+\omega_{1}\omega_{2}^2$,
	$x_3 = \kappa_{1}^2+2\kappa_{2}^2-\gamma^2+t^2-\omega_{2}^2-2\omega_{1}\omega_{2}$ and
	$x_4 = \omega_{1}+2\omega_{2}$.

	We numerically solve the system and find one trajectory satisfying the above Eq.~(\ref{nlcd1}) as shown in Fig.~\ref{NEP5}(a),
	by setting $\gamma>0$, $\kappa_{1}=1$, $\kappa_{2}>0$ and $t>0$ for physical solutions.
	We confirm such a trajectory is the steady solution by
	checking the Lyapunov exponent~\cite{MCT}.
	
	To investigate critical behaviors of nonlinear gain-induced EP (NEP), we include an external perturbation factor $\varepsilon$ onto certain parameters \cite{HOEP2}, \emph{e.g.} $\omega_{1}+\varepsilon$.
	The response of eigenvalues of the nonlinear gained semiclassical Dicke model is shown in Figs.~\ref{NEP5}(b) and (c).
	The solid orange line represents the stable mode and the other dashed lines represent the other four auxiliary modes.
	We note that these five eigenvalues are either real(stable modes) or complex conjugate pairs(auxiliary modes),
	and they become degenerate at NEP(black star).
	They together form the complete basis for nonlinear dynamics~\cite{NEP1,NEP2}.
	Moreover, the eigenvalues $\Delta\omega=\omega-\omega_{s}$  near the NEP are proportional to ${\varepsilon}^{1/5}$ as shown in Fig.~\ref{NEP5}(d), which confirms such EP is the fifth-order NEP (NEP5).

	\subsection{Dynamics evolution}
	
	In this section, we solve for the evolution of light-matter interactions Dicke system described by the Hamiltonians Eq.~(\ref{Eq4}) based on the solutions of the probability amplitudes.
	
	\begin{figure*}[htb]
		\begin{center}
			\centering
			\includegraphics[width=18cm]{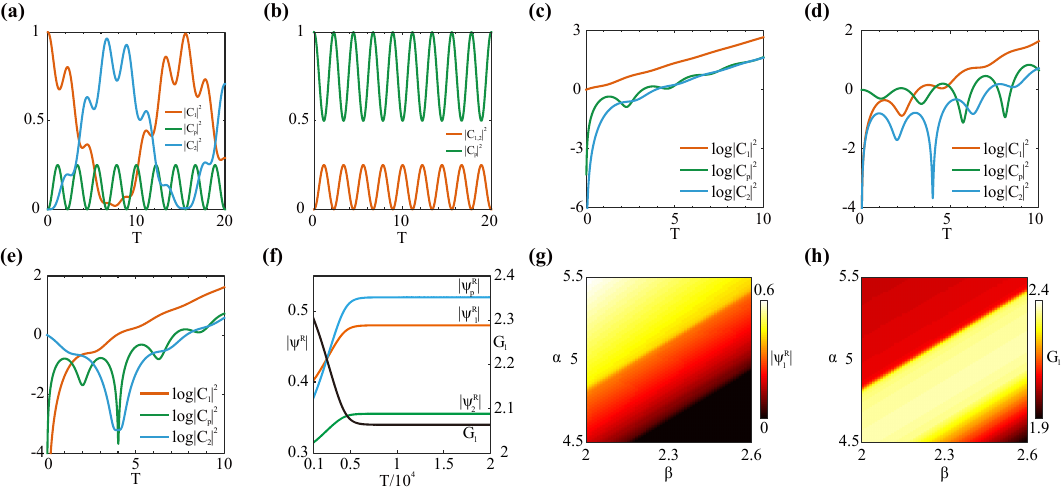}
			\caption{
				(a-b) For Hermitian Dicke model, the probabilities of the $i$-th atom and photon being in states $|\psi_{i}\rangle$ and $|\psi_{p}\rangle$ over Time, with initial states at (a)$|\psi_{1}\rangle$ and (b)$|\psi_{p}\rangle$. Here, we set $\Delta=2$, $\kappa=1$, $\gamma=0$ and $t=0$.
				(c-e) For non-Hermitian Dicke model, the probabilities of the $i$-th atom and photon being in states $|\psi_{i}\rangle$ and $|\psi_{p}\rangle$ over Time, with initial states at (c)$|\psi_{1}\rangle$, (d)$|\psi_{p}\rangle$ and (e)$|\psi_{2}\rangle$. Here, we set $\Delta=2$, $\kappa=1$, $\gamma=0.4$ and $t=0$.
				(f) Evolution of $|\psi^R\rangle=(\psi_1^R, \psi_p^R, \psi_2^R)^T$ and gain $G_{1}(|\psi_{1}^{R}|)$ are shown in colorful lines and black line, respectively.
				After evolving for a long enough time, the nonlinear system will remain in a stable state. Here, $\alpha=5$ and $\beta=2$.
				(g) and (h) Steady states of field amplitude $|\psi_{1}^R|$ and gain $G_{1}(|\psi_{1}^{R}|)$ after evolution for a long enough time as a function of the pumping strength $\alpha$ and the intrinsic losing $\beta$.
				The initial state at $|\psi^{R}(T=0)\rangle=10^{-3}(1,1,1)^T$ and other parameters linked to NEP5 are identical as Figs.~\ref{NEP5}(b-d).
			}
			\label{fig6}
		\end{center}
	\end{figure*}

	Considering the semiclassical version of the non-Hermitian Dicke model and the interaction of two coupled atom ensembles with a single-mode photonic field (Fig.~\ref{fig1}(b)). Let $|\psi_{i}\rangle$ and $|\psi_{p}\rangle$ represent the states of $i$-th atom and photon, $i.e.$, they are the right eigenstates of Hamiltonian $\mathcal{H}$ in in Eq.~(\ref{Eq4}) with corresponding complex eigenvalues plotted in Fig.~\ref{fig2}. The wave function of the non-Hermitian Dicke model can be written in the form
	\begin{eqnarray}
		|\psi(T)\rangle = C_{1}(T)|\psi_{1}\rangle + C_{p}(T)|\psi_{p}\rangle + C_{2}(T)|\psi_{2}\rangle,
	\end{eqnarray}
	where $C_{i}$($i=1, 2$) and $C_{p}$ are the probability amplitudes of finding the atom $i$ in state $|\psi_{i}\rangle$ and the phonon in state $|\psi_{p}\rangle$, respectively. The corresponding time-dependent Schr\"odinger equation is (here we assume $\hbar\equiv1$)
	\begin{eqnarray}
		i\hbar\frac{\partial}{\partial T}|\psi(T)\rangle=\mathcal{H}|\psi(T)\rangle,
	\end{eqnarray}
	The equations of motion for the amplitudes $C_{i}$($i=1, 2$) and $C_{p}$ may be written as
	\begin{subequations}
		\begin{align}
			\dot C_{1} =& (-i\Delta+\gamma)C_{1}-i\frac{\kappa}{\sqrt{2}}C_{p}-itC_{2}, \\
			\dot C_{p} =& -i\frac{\kappa}{\sqrt{2}}(C_{1}+C_{2}), \\
			\dot C_{2} =& (-i\Delta-\gamma)C_{2}-i\frac{\kappa}{\sqrt{2}}C_{p}-itC_{1}.
		\end{align}
	\end{subequations}
	
	Given that the two atomic assembles couplings $t$ is unable to make essential changes for the evolution of wave function, we turn it off for analysis simplification. We first give analytical solutions of the Hermitian Dicke model, and then we present numerical solutions of the non-Hermitian Dicke model, in a given initial state. The solutions of Hermitian Dicke model for $C_{i}$($i=1, 2$) and $C_{p}$ can be written as
	\begin{subequations}
		\begin{align}
			C_{1}(T) =& Pe^{x_{1}T} + Qe^{x_{2}T} - Re^{x_{3}T}, \\
			C_{p}(T) =& -\frac{\sqrt{2}i}{\kappa}(Px_{2}e^{x_{1}T} + Qx_{1}e^{x_{2}T}), \\
			C_{2}(T) =& Pe^{x_{1}T} + Qe^{x_{2}T} + Re^{x_{3}T},
		\end{align}
	\end{subequations}
	where $x_{1} = \frac{i}{2}(\Omega-\Delta)$, $x_{2} = -\frac{i}{2}(\Omega+\Delta)$, $x_{3} = x_{1} + x_{2} = -i\Delta$, and the oscillation frequency is defined as
	\begin{eqnarray}
		\Omega= \sqrt{\Delta^{2}+4\kappa^{2}},
	\end{eqnarray}
	and $P$, $Q$, and $R$ are constants of integration which are determined from the initial conditions:
	\begin{subequations}
		\begin{align}
			P =& \frac{1}{4\Omega}\left [ (\Omega-\Delta)(C_{1}(0)+C_{2}(0))-2\sqrt{2}\kappa C_{p}(0) \right ],\\
			Q =& \frac{1}{4\Omega}\left [ (\Omega+\Delta)(C_{1}(0)+C_{2}(0))+2\sqrt{2}\kappa C_{p}(0) \right ],\\
			R =& \frac{1}{2}(C_{2}(0)-C_{1}(0)),
		\end{align}
	\end{subequations}
	It is not difficult to verify that
	\begin{eqnarray}
		|C_{1}(T)|^2 + |C_{p}(T)|^2 + |C_{2}(T)|^2 = 1,  \label{conservationP}
	\end{eqnarray}
	which is a simple statement of the conservation of probability since the atom is in state $|\psi_{1}\rangle$ or $|\psi_{2}\rangle$ and the photon is in state $|\psi_{p}\rangle$. The conservation of probability is a typical feature of Hermitian systems during the entire evolutionary process.
	
	If we assume that the first atom is initially in the state $|\psi_{1}\rangle$ then $C_{1}(0)=1$, $C_{p}(0)=C_{2}(0)=0$. The probabilities of the $i$-th atom and photon being in states $|\psi_{i}\rangle$ and $|\psi_{p}\rangle$ at time $T$ are given by $|C_{i}(T)|^{2}$ and $|C_{p}(T)|^{2}$.
	
	\begin{subequations}
		\begin{align}
			C_{1,2}(T) =& \frac{1}{2}\left [ \cos\left(\frac{\Omega T}{2}\right) - i\frac{\Delta}{\Omega} \sin\left(\frac{\Omega T}{2}\right) \right ] e^{-i\frac{\Delta T}{2}} \nonumber\\
			& \pm \frac{1}{2} e^{-i\Delta T}, \\
			C_{p}(T) =& -i\frac{\sqrt{2}\kappa}{\Omega} \sin\left(\frac{\Omega T}{2}\right)e^{-i\frac{\Delta T}{2}}, \\
			|C_{1,2}|^2 =& \left [ \cos\left(\frac{\Omega T}{2}\right) \pm \cos\left(\frac{\Delta T}{2}\right) \right ]^2 \nonumber\\
			& + \left [ \frac{\Delta}{\Omega} \sin\left(\frac{\Omega T}{2}\right) \pm \sin\left(\frac{\Delta T}{2}\right) \right ]^2, \\
			|C_{p}|^2 =& \frac{\kappa^2}{\Delta^2 + 4\kappa^2} \left [ 1 - \cos\left(\frac{\Omega T}{2}\right) \right ],
		\end{align}
	\end{subequations}

	If we assume that the photon is initially in the state $|\psi_{p}\rangle$ then $C_{p}(0)=1$, $C_{1}(0)=C_{2}(0)=0$. The probabilities of finding the $i$-th atom or photon are given by
	\begin{subequations}
		\begin{align}
			C_{1}(T) =& C_{2}(T) = -i\frac{\sqrt{2}\kappa}{\Omega}sin(\frac{\Omega T}{2})e^{-i\frac{\Delta T}{2}},\\
			C_{p}(T) =& \left [ cos(\frac{\Omega T}{2}) + i\frac{\Delta}{\Omega}sin(\frac{\Omega T}{2}) \right ]e^{-i\frac{\Delta T}{2}}, \\
			|C_{1}|^{2} =& |C_{2}|^{2} = \frac{\kappa^2}{\Delta^2+4\kappa^2}\left [ 1-cos( \Omega T) \right ], \\
			|C_{p}|^2 =& \frac{\Delta^2+2\kappa^2}{\Delta^2+4\kappa^2}+\frac{2\kappa^2}{\Delta^2+4\kappa^2}cos(\Omega T).
		\end{align}
	\end{subequations}
	We find that the probabilities of the $i$-th atom and photon being in states $|\psi_{i}\rangle$ and $|\psi_{p}\rangle$ both are period oscillation at a frequency $\Omega$.
	
	Next, we introduce non-Hermiticity into our model by activating balanced gain and loss $\gamma$.
	The solutions of non-Hermitian Dicke model for $C_{i}$($i=1, 2$) and $C_{p}$ can be written as
	\begin{subequations}
		\begin{align}
			C_{1}(T) =& -\sum_{i=1}^{3}B_{i}\frac{\sqrt{2}}{\kappa}(\Delta-i\gamma-iy_{i})e^{y_{i}T}, \\
			C_{p}(T) =& -\sum_{i=1}^{3}B_{i}\left [ \frac{2y_{i}^2}{\kappa^2} + \frac{2y_{i}(\gamma+\Delta i)}{\kappa^2} + 1  \right ] e^{y_{i}T}, \\
			C_{2}(T) =& \sum_{i=1}^{3}B_{i}e^{y_{i}T},
		\end{align}
	\end{subequations}
	where $y_{1,2,3}$ are the roots of the cubic equation
	\begin{eqnarray}
		Y^{3}+2\Delta iY^{2}-(\Delta^2+\gamma^2-\kappa^2)Y+\Delta\kappa^2 i=0,
	\end{eqnarray}
	ranked by their real parts, \emph{i.e.} $Re(y_{1})\leq Re(y_{2})\leq Re(y_{3})$.
	The initial conditions determine the constants of integration, $B_{1,2,3}$.
	
	Due to the complexity of solving the cubic equation with complex constant coefficients,
	we numerically solve the time-dependent Schr\"{o}dinger equation for a specified initial state.
	The probabilities of finding the $i$-th atom or photon at time $T$ are shown in Figs.~\ref{fig6}(c-e).
	Their probability amplitudes are increasing exponentially over time due to the maximal real part $Re(y_3)$ being greater than zero.
	Compared to the Hermitian system, a non-Hermitian system tends to violate the law of the conservation of probability in Eq.~(\ref{conservationP}) due to the existence of gain or loss, leading to a non-unitary evolution.
	
	Finally, we consider a nonlinear version of the Dicke model, where the linear gain in Eq.~(\ref{Eq4}) is replaced by a nonlinear saturation gain applied to the first atom.  The time-dependent nonlinear Schr\"{o}dinger equation is given by (here we again assume $\hbar\equiv1$)
	\begin{eqnarray}
		i\hbar\frac{\partial}{\partial T}|\psi^{R}\rangle=\hat{H}_{NL}|\psi^{R}\rangle,
	\end{eqnarray}
	from a nearly zero initial state $|\psi^{R}(T=0)\rangle=10^{-3}(1,1,1)^T$.
	The explicated form of nonlinear Hamiltonian is provided in Eq.~(\ref{Eq8}).
	The colorful lines and black line in Fig.~\ref{fig6}(f) present numerical solutions of $|\psi^{R}|$ and $g(|\psi^{R}_1|)$ by solving the nonlinear Schr\"odinger equation.
	The gain remains at a stable value after evolution for a long enough time, and the final states oscillate at a steady state (see Fig.~\ref{fig6}(f)).
	Furthermore, we investigate the finial steady states as a function of competitive relationships between the pumping strength $\alpha$ and the intrinsic losing $\beta$, presented in Figs.~\ref{fig6}(g) and \ref{fig6}(h).
	When the intrinsic loss dominates, the final states ultimately decay to zero.
	Otherwise, the states oscillate around a stable state.
	
	We conclude that, in Hermitian systems, the conservation of probability stems directly from the Hermiticity of the Hamiltonian, ensuring that the time evolution is unitary and preserves the total probability. However, non-Hermitian systems, which often represent open or dissipative systems, can introduce gain or loss, leading to a non-unitary evolution that typically violates probability conservation.
	
	The introduction of nonlinear mechanisms like saturation gain in non-Hermitian systems can stabilize them. This aligns with the idea that feedback mechanisms may counterbalance the gain or loss, allowing the system to reach steady states after some time. Nonlinear terms often introduce a form of self-regulation, where growth due to gain is counteracted at some threshold, preventing runaway behavior and leading to a dynamic equilibrium.
	
	It would be interesting to explore whether such mechanisms could lead to phenomena like PT-symmetry breaking or stabilization of exceptional points in non-Hermitian systems, potentially providing a richer understanding of the interplay between nonlinearity and dissipation in quantum systems.

	\subsection{1D Dicke chain}
	
\begin{figure}[htb]
		\begin{center}
			\centering
			\includegraphics[width=8.5cm]{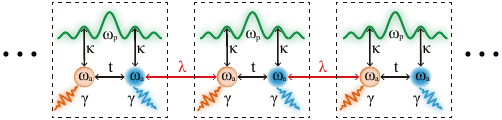}
			\caption{
				The 1D Dicke lattice schematic diagram.
				$\lambda$ denotes the nearest next-neighbor inter-site coupling.
				Other parameters are shown in Fig.~\ref{fig1}(b).
			}
			\label{fig7}
		\end{center}
	\end{figure}

	We investigate topological phase transition and their edge states in 1D non-Hermitian semiclassical  Dicke chain.
	Specifically, we extend the semiclassical Dicke model at Eq.~(\ref{cdicke1}) to 1D chain by introducing the nearest-neighboring inter-site atom-atom coupling $\lambda$, as shown in Fig.~\ref{fig7}.
	The dashed boxes delimit the boundary of the elementary unit cell and other parameters are identical to the model in Fig.~\ref{fig1}(b).
	Thus each intra-site cell contains two coupled atom ensembles interacting with a single-mode photonic field.
	The corresponding Hamiltonian of the system can be expressed as
	\begin{eqnarray}
		\hat{H}^{\rm 1D}_{\rm cDicke} &=&
		\sum_{n} \left [ \sum_{j=1,2}(\omega_a+i\gamma_j)\hat{b}^\dag_{jn}\hat{b}_{jn}+\omega_0\hat{a}_{n}^\dag\hat{a}_{n}+ \right . \\ \nonumber &&t(\hat{b}^\dag_{1n}\hat{b}_{2n}+\hat{b}_{1n}\hat{b}_{2n}^\dag)+\sum_{j=1,2}\frac{\kappa_j}{\sqrt{2}}(\hat{b}_{jn}^\dag\hat{a}_{n}+\\ \nonumber
		&& \left . \hat{b}_{jn}\hat{a}_{n}^\dag) \right ] +\sum_{n}\lambda(\hat{b}_{2n}^\dag\hat{b}_{1,n+1}+\hat{b}_{1n}^\dag\hat{b}_{2,n-1}),
	\end{eqnarray}
	where $\hat{b}^\dag_{jn}$($\hat{a}_{n}^\dag$) and $\hat{b}_{jn}$($\hat{a}_{n}$) are the creation (annihilation) operators for the sublattice $n$.
	In this section, we still keep $\gamma_1=-\gamma_2=\gamma$ as well as taking the frequency of photonic field $\omega_{p}$ as zero energy reference point. After performing Fourier transformations, the Hamiltonian matrix of the system in momentum space can be reexpressed as
	\begin{eqnarray}
		\mathcal{H}_{\rm 1D}(k_x)=
		\begin{bmatrix}
			\Delta+i\gamma & \kappa/\sqrt{2} & t+\lambda e^{-ik_x} \\
			\kappa/\sqrt{2} & 0 & \kappa/\sqrt{2} \\
			t+\lambda e^{ik_x} & \kappa/\sqrt{2} & \Delta-i\gamma
		\end{bmatrix}
		, \label{Ham_of_1D_Dicke_chain}
	\end{eqnarray}
	where $\lambda$ is the inter-site coupling between two nearest-neighboring atoms, and $k_x\in[-\pi, \pi)$ is Bloch vector in the first Brillouin zone.

	To characterize distinct topologically nontrivial properties in gaped bulk energy bands,
	we include the celebrated Zak phase.
	Generally, we define four complex Zak phase components~\cite{NHTI1, NHTI2}
	\begin{eqnarray}
		\gamma_n^{\alpha\beta}=\frac{1}{2\pi}\int_{BZ}-i\langle\psi_n^{\alpha}|\nabla_{k_x}|\psi_n^{\beta}\rangle dk_x,
	\end{eqnarray}
	where $\alpha,\beta=L/R$ denote left/right eigenvectors and $n$ means the energy band index.
	Due to the relation $\gamma_n^{\alpha\beta}=(\gamma_n^{\beta\alpha})^{*}$, we can  obtain real quantized Zak phase in 1D non-Hermitian system by summing over all energy bands below the gap (or Fermi energy level)
	\begin{eqnarray}
		\gamma_n=\sum\limits_{E_n<E_f}\frac{1}{2}(\gamma_n^{LR}+\gamma_n^{RL}). \label{Zakphase}
	\end{eqnarray}
	It should be noted that such quantized Zak phase only owns two discrete values $0$ and $\pi$~\cite{ZakPhase1}, and they sharply switch from $0$ to $\pi$ or vice versa once a topological transformation occurs~\cite{ZakPhase2}.

\begin{figure*}[htb]
		\begin{center}
			\centering
			\includegraphics[width=18cm]{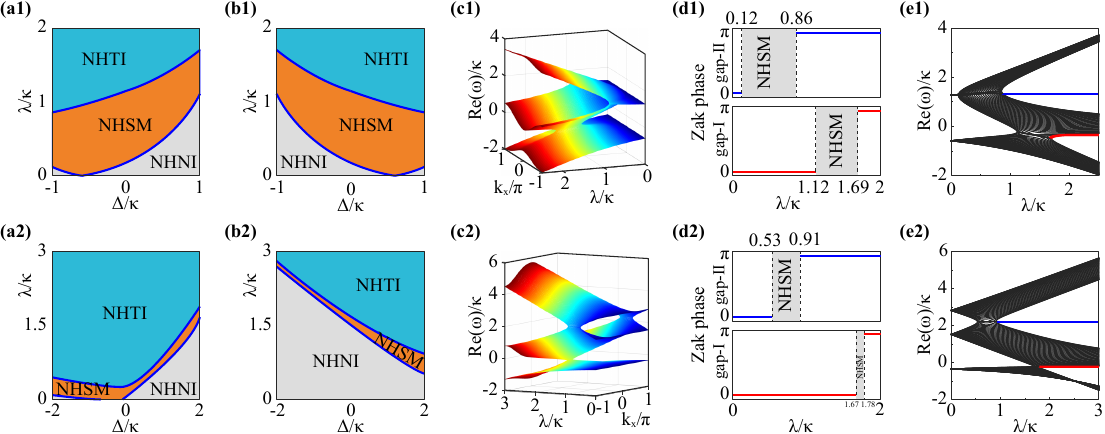}
			\caption{
				(a*) and (b*) Phase diagrams of gap-I and gap-II in non-Hermitian Dicke lattices, where gap-I and gap-II denote the first and second energy bandgaps, respectively.
				The gray region marked as ``NHNI'' represents non-Hermitian normal insulator phase,
				the orange region labeled as ``NHSM'' represents non-Hermitian semimetal phase,
				and the blue region marked as ``NHTI'' represents non-Hermitian topological insulator phase.
				(c*) Bulk energy bands under period boundary conditions.
				(d*) Zak phases of both gap-I and gap-II, which are well-defined when the energy gap is open. In the gray regions labeled ``NHSM'' is ill-defined.
				(e*) Edge states under open boundary conditions for finite chain sites $N = 40$.
				Black lines represent bulk states,
				while red(blue) lines represent topological edge states in the first(second) bandgap.
				The parameters used for panels (*1) are $(\gamma/\kappa, t/\kappa)=(0.5, 0)$, and for panels (*2) $(\gamma/\kappa, t/\kappa)=(0.2, 0.5)$.
				In panels (c1,c2)-(e1,e2), $\Delta/\kappa=1(2)$ and other parameters are identical to those in panels (a1,a2)-(b1,b2).
			}
			\label{fig8}
		\end{center}
	\end{figure*}

	We investigate energy eigenvalues and their topological edge states in 1D non-Hermitian semiclassical Dicke chain at various parameter planes in Figure~\ref{fig8}.
	Given that the two atomic assembles coupling $t$ is unable to make dramatic changes for phase diagrams in both gap and gapless cases,
	we turn it off for demonstration simplification,
	then the model with finite $t$ is presented in Figs~\ref{fig8}(a2)-(e2).
	One can straightforwardly find
	that the Hamiltoninan $\mathcal{H}_{\rm 1D}(k_x)$ at Eq.~(\ref{Ham_of_1D_Dicke_chain})
	owns \rm{PT} symmetry,
	which will reach the criticality under the relation
	$[{\Delta}\mathcal{P}(k_x)+9\kappa^2\mathcal{Q}(k_x)/2]^2=[4\Delta^2-3\mathcal{P}(k_x)][\mathcal{P}^2(k_x)+6\Delta{\mathcal{Q}(k_x)}]$,
	with the wavenumber-dependent coefficients
	$\mathcal{P}(k_x)=\Delta^2+\gamma^2-\kappa^2-(t^2+\lambda^2+2t\lambda\cos{k_x})$
	and
	$\mathcal{Q}(k_x)=\Delta-t-\lambda\cos{k_x}$.
	Thus the PT symmetry breaking occurs
	at either $k_{x}=0$ or $k_{x}=\pi$,
	shown as bottom and upper blue curves in Figs.~\ref{fig8}(a1) and ~\ref{fig8}(b1).
	These two solid curves explicitly divide parameter spaces into three regions, marked as non-Hermitian topological insulators (NHTI), non-Hermitian normal insulators (NHNI), and non-Hermitian semimetals (NHSM), respectively.
	Each region owns distinct topological properties.

	Moreover, it is found that the energy gap of the chain system undergoes opening-closing-recurrent opening transitions by gradually increasing $\lambda$ with given $\Delta$ in Fig.~\ref{fig8}(c1), e.g., $\Delta/\kappa=1$.
	The system is topologically nontrivial(trivial) in the blue(gray) parameter region marked as NHTI(NHNI) with Zak phase obtained from Eq.~(\ref{Zakphase}) to be $\pi$($0$) (see Figs.~\ref{fig8}(a1) and \ref{fig8}(b1)).
	However, the Zak phase in the orange parameter region marked as NHSM is not properly characterized, due to a pair of degeneration points always residing in the bulk energy bands, which can be seen from Fig.~\ref{fig8}(c1).
	Accordingly, Figure~\ref{fig8}(d1) presents the Zak phase switches from zero to $\pi$ with gradually increasing $\lambda$ when $\Delta/\kappa=1$.
	Therefore, they show significantly distinct topological features.
	
	Due to reciprocal couplings in our non-Hermitian 1D Dicke chain, the traditional bulk-boundary correspondence applicable to Hermitian systems remains valid for our non-Hermitian system~\cite{1DNHchain1, 1DNHchain2}.
	From the signal of the bulk Zak phase, one can predict the emergence of edge states~\cite{SSHmodel}.
	For a given set of parameters ($\Delta/\kappa$, $\lambda/\kappa$), the corresponding Zak phase of both gaps can be derived from Eq.~(\ref{Zakphase}).
	Figure~\ref{fig8}(e2) displays the energy spectra of edge states under open boundary conditions with chain sites $N=40$.
	Red(Blue) lines represent a pair of edge states in the first(second) bandgap (denoted henceforth as ``gap-I'' and ``gap-II''),
	while black lines indicate trivial bulk states.
	These results agree well with the predictions of the Zak phase,
	where the NHTI(NHNI) regions correspond to Zak phases of $\pi$($0$),
	indicating the presence(absence) of edge states in the gaps.
	Similar results for the model with finite $t$ are illustrated in Figs.~\ref{fig8}(a2)-(e2).

	\section{Non-Hermitian Quantum Dicke model} \label{QuantumDickeModel}
	
	
	\subsection{Quantum signature of EPs}
	We investigate exceptional points in the quantum regime of the Dicke model by applying the  Lindblad master equation~\cite{manzano2020aip}, i.e.
	\begin{eqnarray}~\label{lme1}
		\frac{d}{dt}\hat{\rho}_s &=& -i[\hat{H}_{\rm Dicke},\hat{\rho}_s]+ \\ &&
		\sum_j\frac{\gamma_j}{N_j}[2\hat{J}^-_j\hat{\rho}_s\hat{J}^+_j-\hat{J}^+_j\hat{J}^-_j\hat{\rho}_s-\hat{\rho}_s\hat{J}^+_j\hat{J}^-_j], \nonumber
	\end{eqnarray}
	where the dissipation of photons is assumed to be much weaker than the atom counterpart.
	If we naively ignore the quantum jump processes
	($\hat{J}^-_j\hat{\rho}_s\hat{J}^+_j$),
	the dissipative dynamics is reduced to
	$ \frac{d}{dt}\hat{\rho}_s{\approx}-i(\hat{H}^\dag_{\rm eff}\hat{\rho}_s-\hat{\rho}_s\hat{H}_{\rm eff})$,
	with the effective non-Hermitian Dicke Hamiltonian
	$\hat{H}_{\rm eff}=\hat{H}_{\rm Dicke}
	+i\sum_j\gamma_j/N_j\hat{J}^+_j\hat{J}^-_j$.
	In particular, the single qubit case (i.e. $N_j=1$) is the generic quantum light-matter interacting model (i.e. Jaynes-Cummings model).
	We should admit that it is rather difficult to analytically obtain exceptional points with small photon excitation even at $N_j=1$.
	While at large $n$ regime,
	under the basis
	$\{|\uparrow\uparrow{n}{\rangle}, |\uparrow\downarrow{n+1}{\rangle}, |\downarrow\uparrow{n+1}{\rangle}, |\downarrow\downarrow{n+2}{\rangle}\}$
	the non-Hermitian Hamiltonian can be approximately simplified as
	$\hat{H}_{\rm eff}{\approx}[\omega_p(n+1)-i\frac{\Gamma}{2}]\mathrm{I}_{4\times{4}}+\mathcal{H}$,
	where the kernel component denotes
	\begin{equation}~\label{lme1}
		\mathcal{H}=
		\begin{bmatrix}
			\delta-i\Gamma & \kappa\sqrt{n}/\sqrt{2} &  \kappa\sqrt{n}/\sqrt{2} & 0 \\
			\kappa\sqrt{n}/\sqrt{2} & i\gamma & t & \kappa\sqrt{n}/\sqrt{2} \\
			\kappa\sqrt{n}/\sqrt{2} & t & -i\gamma & \kappa\sqrt{n}/\sqrt{2} \\
			0 & \kappa\sqrt{n}/2 & \kappa\sqrt{n}/\sqrt{2} & -\delta+i\Gamma
		\end{bmatrix},
	\end{equation}
	with
	$\delta=\omega_a-\omega_p$,
	$\Gamma=(\gamma_2+\gamma_1)/2$,
	and
	$\gamma=(\gamma_2-\gamma_1)/2$.
	If we further consider the absence of atom-atom coupling $t=0$
	and asymmetric dissipation condition of two atoms
	$\gamma_2{\ll}\gamma_1$ (if we set $\gamma_1{\ll}\gamma_2$ we get the similar result),
	which leads to $\gamma=-\Gamma$.
	Thus the positions of EP2 under  $\delta=0$ can be determined at
	\begin{eqnarray}
		\Gamma={\pm} \kappa\sqrt{n}/\sqrt{2},
	\end{eqnarray}
	with the expression of eigenvalues
	$E^2_{\pm,n}=-(\Gamma^2-{n}\kappa^2){\pm}\sqrt{(\Gamma^2-{n}\kappa^2)^2-\Gamma^4}$
	and $n{\ge}0$.
	We note that the exceptional points here highly rely on the photon quantized number $n$, which is the signature of the quantum effect.

	\begin{figure}[htb]
		\begin{center}
			\centering
			\includegraphics[width=8.5cm]{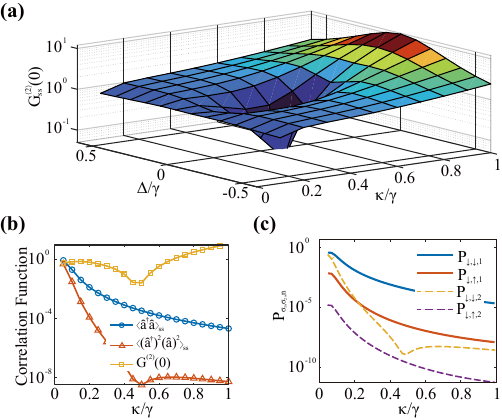}
			\caption{
				(a) Zero-time-delay correlation function $G^{(2)}_{\rm ss}(0)$ at steady state by tuning energy bias $\Delta$ and qubit-photon coupling strength $\kappa$.
				(b) The correlation function components and
				(c) local steady-state populations $P_{\sigma_1,\sigma_2,n}$ with $\sigma_j=\uparrow,\downarrow$ and $n=1,2$, at $\Delta=0$. Other system parameters are given by $t=0$, $\gamma_1=1$, $\gamma_2=0.3$, and $\eta=0.01$.
			}
			\label{fig:G2s}
		\end{center}
	\end{figure}

	In contrast, if we consider the single excitation case of the quantum Dicke model.
	Under the basis
	$\{|\uparrow\downarrow{0}{\rangle}, |\downarrow\uparrow{0}{\rangle}, |\downarrow\downarrow{1}{\rangle}\}$,
	the reduced effective Hamiltonian is described as
	$\hat{H}_{\rm eff}{=}\omega_0\mathrm{I}_{3\times{3}}+\mathcal{H}$,
	with
	\begin{equation}
		\mathcal{H}=
		\begin{bmatrix}
			-i\gamma_1 & t & \kappa/\sqrt{2} \\
			t & -i\gamma_2 & \kappa/\sqrt{2}\\
			\kappa/\sqrt{2} & \kappa/\sqrt{2} & -\delta
		\end{bmatrix}.
	\end{equation}
	Hence the EP2 solution is generally expected to be observed, for the kernel Hamiltonian component is analogous with the non-Hermitian semiclassical Dicke model in the absence of atom gain.
	In particular, once we set $\gamma_2=-\gamma_1=\gamma$, EP3 can also be detected.
	However, these are quantum signatures of exceptional points under the Hilbert space.

	\subsection{Photon statistics}
	
	The nonclassicality of photons is widely regarded as a crucial topic in the quantum community,
	which attracts persistent attention in quantum optics
	and quantum information processing.
	The two-photon correlation function is considered as one main characteristics, which was initially proposed by R. J. Glauber in quantum theory of optical coherence.
	Consequently, two-photon correlation function has been extensively used in various quantum light-matter interaction systems,
	including dissipative quantum Rabi and finite-size Dicke models.
	The steady-state two-photon correlation function with no time delay is defined as
	\begin{eqnarray}
		G^{(2)}_{\rm ss}(\tau)=\frac{{\langle}\hat{a}^\dag\hat{a}^\dag(\tau)\hat{a}(\tau)\hat{a}{\rangle}_{\rm ss}}{{\langle}\hat{a}^\dag\hat{a}{\rangle}^2_{\rm ss}},
	\end{eqnarray}
	where ${\langle}\hat{O}{\rangle}_{\rm ss}
	=\textrm{Tr}\{\hat{O}\hat{\rho}_{\rm ss}\}$,
	with $\hat{\rho}_{\rm ss}$ the steady state of the quantum system.

	We investigate the influence of \rm{PT} symmetry on two-photon statistics.
	It is known that the steady state at dynamical equation Eq.~(\ref{lme1}) is the trivial ground state of $\hat{D}_{\rm Dicke}$, i.e. $\hat{\rho}_{s}=|\downarrow\downarrow{0}{\rangle}{\langle}\downarrow\downarrow{0}|$.
	Thus, to efficiently observe the nonclassical photon correlation,
	we include a weak driving field to excite photons,
	leading to the system Hamiltonian
	$\hat{H}_{\rm t-Dicke}=\hat{H}_{\rm Dicke}+\eta(e^{i\omega_dt}\hat{a}+e^{-i\omega_dt}\hat{a}^\dag)$,
	with $\eta$ and $\omega_d$ being driving amplitude and frequency, respectively.
	Under the rotating framework, i.e.
	the rotated Hamiltonian becomes
	\begin{eqnarray}
		\hat{H}_{\rm R}&=&
		\sum_{j=1,2}\Delta_j\hat{J}^z_j
		+\Delta_p\hat{a}^\dag\hat{a}
		+\frac{t}{\sqrt{N_1N_2}}(\hat{J}^+_1\hat{J}^-_2+\hat{J}^-_1\hat{J}^+_2)\nonumber\\
		&&+\sum_{j=1,2}\frac{\kappa_j}{\sqrt{2N_j}}(\hat{a}^\dag\hat{J}^-_j+h.c.)
		+\eta(\hat{a}+\hat{a}^\dag),
	\end{eqnarray}
	where the detuning frequencies denote
	$\Delta_j=\omega_j-\omega_d$
	and
	$\Delta_p=\omega_p-\omega_d$.
	Consequently, the Lindblad master equation after rotating transformation becomes
	$\frac{d}{dt}\hat{\rho}_s=-i[\hat{H}_{\rm R},\hat{\rho}_s]+\sum_j{\gamma_j}/{N_j}(2\hat{J}^-_j\hat{\rho}_s\hat{J}^+_j-\hat{J}^+_j\hat{J}^-_j\hat{\rho}_s-\hat{\rho}_s\hat{J}^+_j\hat{J}^-_j)$.
	
	We investigate steady-state two-photon correlation function with weak coherent photon driving (e.g., $\eta=0.01\gamma$ in Fig.~\ref{fig:G2s}).
	Under weak system-reservoir interactions, it is found that the steady-state solution can be approximately obtained by the dynamical equation
	\begin{eqnarray}
		\frac{d}{dt}\hat{\rho}_s{\approx}i[\hat{\rho}_s\hat{H}_{\rm NH}-\hat{H}^\dag_{\rm NH}\hat{\rho}_s],~\label{qnheq}
	\end{eqnarray}
	with $\hat{H}_{\rm NH}=\hat{H}_{\rm R}
	+i\sum_j{\gamma_j}/{N_j}\hat{J}^+_j\hat{J}^-_j$,
	which is also numerically confirmed.
	This leads to the steady state $\hat{\rho}_{\rm ss}{\approx}|\psi_{\rm ss}{\rangle}{\langle}\psi_{\rm ss}|$,
	where $|\psi_{\rm ss}{\rangle}$ is the eigenvector of $\hat{H}_{\rm NH}$ with the corresponding eigenvalue owning the minimal imaginary component.
	
	Fig.~\ref{fig:G2s}(a) exhibits the steady-state correlation function with no time delay, based on the non-Hermitian dynamical equation Eq.~(\ref{qnheq}).
	Near the resonant regime (i.e. $\Delta{\approx}0$) the antibunching-to-bunching transition is clearly shown with the increase of qubit-photon coupling strength $\kappa$.
	Such nonclassical transition of photons is dramatically suppressed as the energy bias increases.
	
	Then, we analyze two components of the two-photon correlation function, i.e. one-photon and two-photon terms in Fig.~\ref{fig:G2s}(b),
	both showing a monotonic decrease by enhancing the coupling strength $\kappa$ before $\kappa{\lesssim}0.5\gamma$.
	However, the two-photon term ${\langle}(\hat{a}^\dag)^2(\hat{a})^2{\rangle}_{\rm ss}$
	shows comparative enhancement by further increasing $\kappa$.
	Furthermore, based on the condition of weak photon excitations and local population distributions in Fig.~\ref{fig:G2s}(c),
	one-photon and two-photon correlation terms can be approximately expressed as
	${\langle}\hat{a}^\dag\hat{a}{\rangle}_{\rm ss}{\approx}
	(P_{\downarrow,\downarrow,1}+2P_{\downarrow,\downarrow,2})$
	and
	${\langle}(\hat{a}^\dag)^2(\hat{a})^2{\rangle}_{\rm ss}
	{\approx}2P_{\downarrow,\downarrow,2}$,
	with the population
	$P_{\sigma_1,\sigma_2,n}=
	{\langle}\sigma_1,\sigma_2,n|\hat{\rho}_{\rm ss}|\sigma_1,\sigma_2,n{\rangle}$
	under the local basis $\{|\sigma_1,\sigma_2,n{\rangle}\}$.
	Hence, the two-photon correlation function can be approximately expressed as
	$G^{(2)}(0){\approx}2P_{\downarrow,\downarrow,2}/(P_{\downarrow,\downarrow,1})^2$.
	The dip structure of $P_{\downarrow,\downarrow,2}$
	by tuning up $\kappa$ is crucial in realizing the antibunching-to-bunching nonclassical transition of photons.

	\section{Relevant Physical Systems} \label{PossibleRealizations}
	
	The Dicke model has been experimentally realized in different setups. The semiclassical Dicke model can be mapped into a model consisting of three interacting resonator modes with different dissipation~\cite{HOEP7}. This classical regime can be realized in optics (acoustics) where all  three resonator are photonic (acoustic) modes. There are a number of ways to tune the frequency and loss (gain) of each resonator in photonics. One of the options is to use the thermo-optic effect~\cite{pQTP}. Gain and loss can be controlled by, e.g., doping rare earth elements as the gain medium, or tuning the dissipation. In acoustics, the gain can be realized through electrically assisted thermoacoustic generation of sound, as shown in Ref.~\cite{NHWisper}. In acoustics, the detuning and the couplings between the resonators can be well-controlled by the geometry of the acoustic structures (both the resonators and their connections). In genuine light-matter interacting systems, the non-Hermitian Dicke model can be realized by coupling a photonic microcavity mode with excitons in two quantum wells where the excitonic gain and loss in each quantum well can be controlled electrically. The second-order and third-order EPs in our model can then be achieved by controlling the detuning and the couplings in these setups.
	
	For the non-Hermitian Dicke model in the quantum regime, it may be established in a setup where two cavities are connected via a fine optical fiber, each containing trapped ions, i.e., qubits~\cite{hpu2019prl}.
	The hybrid system of two cavities interacting with the common fiber has one dominant mode closing to the qubit transition frequency~\cite{hjc2007pra},
	whereas other modes could be ignored.
	Moreover, an external field together with the dominant photon mode drives a Raman transition between two low-energy states of ions, marked as $|0{\rangle}$
	and $|1{\rangle}$.
	Therefore, the Dicke model is described as
	\begin{eqnarray}
		\hat{H}_{\rm D}=\omega\hat{a}^\dag\hat{a}+\sum^N_{j=1}[\frac{\varepsilon}{2}\hat{\sigma}^j_z+\frac{g}{\sqrt{N}}\hat{\sigma}^j_x(\hat{a}^\dag+\hat{a})],
	\end{eqnarray}
	where $\hat{a}^\dag~(\hat{a})$ denotes
	the creating(annihilating) operator of the dominant photon mode with the frequency $\omega$,
	$\hat{\sigma}_\alpha~(\alpha=x,y,z)$ are the Pauli operators of trapped ions, composed of  two states
	$|0{\rangle}$
	and $|1{\rangle}$,
	$\varepsilon$ is the splitting energy of ions, $g$ is the ion-photon coupling strength, and $N$ is the number of ions.
	Moreover, we consider the main source of dissipation as spontaneous
	emission of ions,
	which leads to Eq.~(\ref{lme1}).

	\section{Conclusion and discussions} \label{Summary}
	
	In this work, we propose a generalized Dicke model and study the non-Hermitian effects of such a model. In the semiclassical limit, we uncover the emergence of higher-order EPs, which substantially affect the system's behavior and enhance its sensitivity to external perturbations. By incorporating the nonlinear saturation gain, higher-order nonlinear EPs can emerge at certain parameters where the non-Hermitian eigenstates regain completeness in the temporal dynamics. Moreover, rich topological phase transitions are found where the non-Hermitian effects play an important role. Non-Hermitian semimetal phases are found where the EPs are extended to exceptional lines.
	
	For the quantum non-Hermitian Dicke model, we investigate the quantum characteristics of the EPs as well as the transition between the photon anti-bunching behavior and the photon bunching behavior in the steady states, which can be effectively described by the non-Hermitian dynamics. The simplicity of our non-Hermitian Dicke model allows for its implementation in various light-matter interaction systems.
	
	Moreover, the non-Hermitian effects in our model offer potential advantages in applications like enhanced quantum sensing, robust state transfer, and precision control over topological phase transitions.
	Our findings provide insights into non-Hermitian physics in light-matter interacting systems and open a pathway for advancing photonic and quantum technologies with non-Hermitian effects.
	
	Non-Hermitian Hamiltonians often correspond to the semiclassical limit of open quantum systems. In the quantum regime, in contrast, theoretical descriptions of open quantum systems are usually based on quantum master equations, the path-integral approach~\cite{pathintegral}, and the nonequilibrium Green's function method~\cite{koch,phrep}. Nevertheless, non-Hermitian Hamiltonians are able to give much insight into the underlying physics, demonstrating their value in predicting many novel phenomena of nonequilibrium systems. The process of going from the quantum regime to the semiclassical limit is often not explicit in many works. Our model and theory here serve as an example to analyze both the quantum and the semiclassical regimes of  non-Hermitian physics.

	\section*{APPENDIX A: CONSTRUCTION OF EP3}
	
	We start from a semiclassical non-Hermitian Dicke model,
	as shown in Fig.~\ref{fig1}(a),
	by introducing gain or loss breaking the hermiticity,
	\begin{eqnarray}
		\hat{H}=
		\begin{bmatrix}
			\Delta + i\gamma_{1} & \kappa/\sqrt{2} & t\\
			\kappa/\sqrt{2}  & 0 &\kappa/\sqrt{2} \\
			t & \kappa/\sqrt{2} & \Delta + i\gamma_{2}
		\end{bmatrix}, \label{DickeSupp}
	\end{eqnarray}
	where positive(negative) $\gamma_{i}$ indicating gain(loss),
	and other parameters as same as Eq.~(\ref{Eq4}) in the main text.
	We solve eigenfunction to obtian eigenvalues of Hamiltonian in Eq.~(\ref{DickeSupp}),
	\begin{eqnarray}
		\mathrm{Det}(\hat{H}-\omega\hat{I})=0. \label{CharacterPoly}
	\end{eqnarray}
	Here we firstly focus on the three multiple roots $\omega_{EP3}$ of above eigenfunction,
	\begin{eqnarray}
		(\omega-\omega_{EP3})^3=0.
		\label{threeMultiple}
	\end{eqnarray}
	We split the real and imaginary parts of Eq.~(\ref{CharacterPoly}) in the left hand. The imaginary part is
	\begin{eqnarray}
		(\gamma_{1}+\gamma_{2})[\kappa^2/2+\omega(\Delta-\omega)]=0, \label{ImagPart}
	\end{eqnarray}
	and the real part is a third-order characteristic polynomial regarding eigenfrequency $\omega$,
	\begin{eqnarray}
		\omega^3-2\Delta\omega^2+(\Delta^2-\kappa^2-\gamma_{1}\gamma_{2}-t^2)\omega+\kappa^2(\Delta-t)=0. \label{RealPart}
	\end{eqnarray}
	Comparing the coefficients in Eq.~(\ref{threeMultiple}) and Eq.~(\ref{RealPart}), we obtain
	\begin{subequations}
		\begin{eqnarray}
			\Delta=\frac{3}{2}\omega_{EP3},\\ \label{EP3Sb}
			\gamma_{1}\gamma_{2}=-\frac{1}{\kappa^4}(\omega_{EP3}^2+\kappa^2)^3,\\
			t=\frac{3}{2}\omega_{EP3}+\frac{\omega_{EP3}^3}{\kappa^2},
		\end{eqnarray}
	\end{subequations}
	From the Eq.~(\ref{EP3Sb}), the values of $\gamma_{1}$ and $\gamma_{2}$ should embrace opposite signs if existence of EP3 in above Hamiltonian of Eq.~(\ref{DickeSupp}).
	There are no EP3s if only introducing loss (or gain) in non-Hermitian Dicke model.
	Thus, in the sake of simplification,
	we set $\gamma_{1}=-\gamma_{2}=\gamma>0$ in our non-Hermitian Dicke model,
	and the Eq.~(\ref{ImagPart}) can be satisfied automatically.
	Finally, the trajectory of EP3 can be expressed as the parameter equation regarding eigenfrequencies of EP3,
	\begin{subequations}
		\begin{eqnarray}
			\Delta = \frac{3}{2}\omega_{EP3},\\
			\gamma = \frac{1}{\kappa^2}(\omega_{EP3}^2+\kappa^2)^\frac{3}{2}, \\
			t=\frac{3}{2}\omega_{EP3}+\frac{\omega_{EP3}^3}{\kappa^2}.
		\end{eqnarray} \label{EP3SS}
	\end{subequations}
	
	Since we found that the necessary condition for the emergence of EP3 is that the system has balance loss and gain,
	\emph{e.g.} $\gamma_{1}=-\gamma_{2}=\gamma$,
	next we continue to analysis EP2 in a similar way.
	The system emerges a EP2, which implies its eigenvalues spectra $\omega_{EP2}$ with algebraic multiplicity to be 2 and another single eigenvalue $\omega_{3}$.
	The corresponding cubic characteristic equation now can be expressed as
	\begin{eqnarray}
		(\omega-\omega_{EP2})^2(\omega-\omega_{3})=0. \label{CubicEP2}
	\end{eqnarray}
	Comparing the coefficients in Eq.~(\ref{RealPart}) and Eq.~(\ref{CubicEP2}) (or according to Vieta's formulas for cubic equation), we obtain
	\begin{subequations}
		\begin{eqnarray}
			\Delta=\frac{t\kappa^2+2\omega_{EP2}^3}{\kappa^2+2\omega_{EP2}^2}, \\
			\gamma=\frac{\sqrt{\kappa^2+\omega_{EP2}^2}(\kappa^2+2t\omega_{EP2})}{\kappa^2+2\omega_{EP2}^2}, \\
			\omega_{3}=\frac{2(t-\omega_{EP2})\kappa^2}{\kappa^2+2\omega_{EP2}^2}. \label{EP2Sc}
		\end{eqnarray} \label{EP2SS}
	\end{subequations}
	
	On the one hand, we notice that,
	if $\omega_{EP2}=\omega_{3}$, the Eq.~(\ref{EP2SS}) can be reduced to Eq.~(\ref{EP3SS}).
	On the other hand, the value of $\omega_{EP2}$ can be greater than $\omega_{3}$ only if $t/\gamma < 1$, which can be seen from Eq.~(\ref{EP2Sc}).
	Therefore, the yellow curve $t/\gamma = 1$ in Fig.~\ref{fig1}(b) is a critical middle process which transfer a EP2 between two adjacent energy gaps.
	Below the yellow line in the region of class-I,
	two EP2 appear in the same energy gap while above the yellow line in the region of class-II,
	two EP2 appear in adjacent energy gaps and the according EPFP is shown in Fig.~\ref{fig2}(d).

	\bibliography{reference}

\begin{thebibliography}{80}%
\makeatletter
\providecommand \@ifxundefined [1]{%
 \@ifx{#1\undefined}
}%
\providecommand \@ifnum [1]{%
 \ifnum #1\expandafter \@firstoftwo
 \else \expandafter \@secondoftwo
 \fi
}%
\providecommand \@ifx [1]{%
 \ifx #1\expandafter \@firstoftwo
 \else \expandafter \@secondoftwo
 \fi
}%
\providecommand \natexlab [1]{#1}%
\providecommand \enquote  [1]{``#1''}%
\providecommand \bibnamefont  [1]{#1}%
\providecommand \bibfnamefont [1]{#1}%
\providecommand \citenamefont [1]{#1}%
\providecommand \href@noop [0]{\@secondoftwo}%
\providecommand \href [0]{\begingroup \@sanitize@url \@href}%
\providecommand \@href[1]{\@@startlink{#1}\@@href}%
\providecommand \@@href[1]{\endgroup#1\@@endlink}%
\providecommand \@sanitize@url [0]{\catcode `\\12\catcode `\$12\catcode
  `\&12\catcode `\#12\catcode `\^12\catcode `\_12\catcode `\%12\relax}%
\providecommand \@@startlink[1]{}%
\providecommand \@@endlink[0]{}%
\providecommand \url  [0]{\begingroup\@sanitize@url \@url }%
\providecommand \@url [1]{\endgroup\@href {#1}{\urlprefix }}%
\providecommand \urlprefix  [0]{URL }%
\providecommand \Eprint [0]{\href }%
\providecommand \doibase [0]{http://dx.doi.org/}%
\providecommand \selectlanguage [0]{\@gobble}%
\providecommand \bibinfo  [0]{\@secondoftwo}%
\providecommand \bibfield  [0]{\@secondoftwo}%
\providecommand \translation [1]{[#1]}%
\providecommand \BibitemOpen [0]{}%
\providecommand \bibitemStop [0]{}%
\providecommand \bibitemNoStop [0]{.\EOS\space}%
\providecommand \EOS [0]{\spacefactor3000\relax}%
\providecommand \BibitemShut  [1]{\csname bibitem#1\endcsname}%
\let\auto@bib@innerbib\@empty
\bibitem [{\citenamefont {Ding}\ \emph {et~al.}(2022)\citenamefont {Ding},
  \citenamefont {Fang},\ and\ \citenamefont {Ma}}]{gcma2022nrp}%
  \BibitemOpen
  \bibfield  {author} {\bibinfo {author} {\bibfnamefont {K.}~\bibnamefont
  {Ding}}, \bibinfo {author} {\bibfnamefont {C.}~\bibnamefont {Fang}}, \ and\
  \bibinfo {author} {\bibfnamefont {G.}~\bibnamefont {Ma}},\ }\bibfield
  {title} {\enquote {\bibinfo {title} {Non-hermitian topology and
  exceptional-point geometries},}\ }\href {\doibase 10.1038/s42254-022-00516-5}
  {\bibfield  {journal} {\bibinfo  {journal} {Nat. Rev. Phys.}\ }\textbf
  {\bibinfo {volume} {4}},\ \bibinfo {pages} {745} (\bibinfo {year}
  {2022})}\BibitemShut {NoStop}%
\bibitem [{\citenamefont {Zhu}\ \emph {et~al.}(2023)\citenamefont {Zhu},
  \citenamefont {Deng}, \citenamefont {Liu}, \citenamefont {Lu}, \citenamefont
  {Wang}, \citenamefont {Lin}, \citenamefont {Huang}, \citenamefont {Jiang},\
  and\ \citenamefont {Liu}}]{zhu2023}%
  \BibitemOpen
  \bibfield  {author} {\bibinfo {author} {\bibfnamefont {W.}~\bibnamefont
  {Zhu}}, \bibinfo {author} {\bibfnamefont {W.}~\bibnamefont {Deng}}, \bibinfo
  {author} {\bibfnamefont {Y.}~\bibnamefont {Liu}}, \bibinfo {author}
  {\bibfnamefont {J.}~\bibnamefont {Lu}}, \bibinfo {author} {\bibfnamefont
  {H.-X.}\ \bibnamefont {Wang}}, \bibinfo {author} {\bibfnamefont {Z.-K.}\
  \bibnamefont {Lin}}, \bibinfo {author} {\bibfnamefont {X.}~\bibnamefont
  {Huang}}, \bibinfo {author} {\bibfnamefont {J.-H.}\ \bibnamefont {Jiang}}, \
  and\ \bibinfo {author} {\bibfnamefont {Z.}~\bibnamefont {Liu}},\ }\bibfield
  {title} {\enquote {\bibinfo {title} {Topological phononic metamaterials},}\
  }\href {\doibase 10.1088/1361-6633/aceeee} {\bibfield  {journal} {\bibinfo
  {journal} {Rep. Prog. Phys.}\ }\textbf {\bibinfo {volume} {86}},\ \bibinfo
  {pages} {106501} (\bibinfo {year} {2023})}\BibitemShut {NoStop}%
\bibitem [{\citenamefont {Lin}\ \emph {et~al.}(2023)\citenamefont {Lin},
  \citenamefont {Tai}, \citenamefont {Li},\ and\ \citenamefont
  {Lee}}]{front_lee}%
  \BibitemOpen
  \bibfield  {author} {\bibinfo {author} {\bibfnamefont {R.}~\bibnamefont
  {Lin}}, \bibinfo {author} {\bibfnamefont {T.}~\bibnamefont {Tai}}, \bibinfo
  {author} {\bibfnamefont {L.}~\bibnamefont {Li}}, \ and\ \bibinfo {author}
  {\bibfnamefont {C.~H.}\ \bibnamefont {Lee}},\ }\bibfield  {title} {\enquote
  {\bibinfo {title} {Topological non-hermitian skin effect},}\ }\href {\doibase
  10.1007/s11467-023-1309-z} {\bibfield  {journal} {\bibinfo  {journal} {Front.
  Phys.}\ }\textbf {\bibinfo {volume} {18}},\ \bibinfo {pages} {53605}
  (\bibinfo {year} {2023})}\BibitemShut {NoStop}%
\bibitem [{\citenamefont {Bergholtz}\ \emph {et~al.}(2021)\citenamefont
  {Bergholtz}, \citenamefont {Budich},\ and\ \citenamefont {Kunst}}]{RMP_Emil}%
  \BibitemOpen
  \bibfield  {author} {\bibinfo {author} {\bibfnamefont {E.~J.}\ \bibnamefont
  {Bergholtz}}, \bibinfo {author} {\bibfnamefont {J.~C.}\ \bibnamefont
  {Budich}}, \ and\ \bibinfo {author} {\bibfnamefont {F.~K.}\ \bibnamefont
  {Kunst}},\ }\bibfield  {title} {\enquote {\bibinfo {title} {Exceptional
  topology of non-hermitian systems},}\ }\href {\doibase
  10.1103/RevModPhys.93.015005} {\bibfield  {journal} {\bibinfo  {journal}
  {Rev. Mod. Phys.}\ }\textbf {\bibinfo {volume} {93}},\ \bibinfo {pages}
  {015005} (\bibinfo {year} {2021})}\BibitemShut {NoStop}%
\bibitem [{\citenamefont {Ashida}\ \emph
  {et~al.}(2020{\natexlab{a}})\citenamefont {Ashida}, \citenamefont {Gong},\
  and\ \citenamefont {Ueda}}]{rev_Ueda}%
  \BibitemOpen
  \bibfield  {author} {\bibinfo {author} {\bibfnamefont {Y.}~\bibnamefont
  {Ashida}}, \bibinfo {author} {\bibfnamefont {Z.}~\bibnamefont {Gong}}, \ and\
  \bibinfo {author} {\bibfnamefont {M.}~\bibnamefont {Ueda}},\ }\bibfield
  {title} {\enquote {\bibinfo {title} {Non-hermitian physics},}\ }\href
  {\doibase 10.1080/00018732.2021.1876991} {\bibfield  {journal} {\bibinfo
  {journal} {Adv. Phys.}\ }\textbf {\bibinfo {volume} {69}},\ \bibinfo {pages}
  {249} (\bibinfo {year} {2020}{\natexlab{a}})}\BibitemShut {NoStop}%
\bibitem [{\citenamefont {Okuma}\ and\ \citenamefont
  {Saito}(2023)}]{rev_Saito}%
  \BibitemOpen
  \bibfield  {author} {\bibinfo {author} {\bibfnamefont {N.}~\bibnamefont
  {Okuma}}\ and\ \bibinfo {author} {\bibfnamefont {M.}~\bibnamefont {Saito}},\
  }\bibfield  {title} {\enquote {\bibinfo {title} {Non-hermitian topological
  phenomena: A review},}\ }\href {\doibase
  10.1146/annurev-conmatphys-040521-033133} {\bibfield  {journal} {\bibinfo
  {journal} {Ann. Rev. Cond. Matt. Phys.}\ }\textbf {\bibinfo {volume} {14}},\
  \bibinfo {pages} {83} (\bibinfo {year} {2023})}\BibitemShut {NoStop}%
\bibitem [{\citenamefont {Wang}\ \emph {et~al.}(2023)\citenamefont {Wang},
  \citenamefont {Fu}, \citenamefont {Mao}, \citenamefont {Qie}, \citenamefont
  {Stone},\ and\ \citenamefont {Yang}}]{cqwang2023aop}%
  \BibitemOpen
  \bibfield  {author} {\bibinfo {author} {\bibfnamefont {C.}~\bibnamefont
  {Wang}}, \bibinfo {author} {\bibfnamefont {Z.}~\bibnamefont {Fu}}, \bibinfo
  {author} {\bibfnamefont {W.}~\bibnamefont {Mao}}, \bibinfo {author}
  {\bibfnamefont {J.}~\bibnamefont {Qie}}, \bibinfo {author} {\bibfnamefont
  {A.~D.}\ \bibnamefont {Stone}}, \ and\ \bibinfo {author} {\bibfnamefont
  {L.}~\bibnamefont {Yang}},\ }\bibfield  {title} {\enquote {\bibinfo {title}
  {Non-hermitian optics and photonics: from classical to quantum},}\ }\href
  {\doibase 10.1364/AOP.475477} {\bibfield  {journal} {\bibinfo  {journal}
  {Adv. Opt. Photon.}\ }\textbf {\bibinfo {volume} {15}},\ \bibinfo {pages}
  {442} (\bibinfo {year} {2023})}\BibitemShut {NoStop}%
\bibitem [{\citenamefont {Meng}\ \emph {et~al.}(2024)\citenamefont {Meng},
  \citenamefont {Ang},\ and\ \citenamefont {Lee}}]{chlee2024apl}%
  \BibitemOpen
  \bibfield  {author} {\bibinfo {author} {\bibfnamefont {H.}~\bibnamefont
  {Meng}}, \bibinfo {author} {\bibfnamefont {Y.~S.}\ \bibnamefont {Ang}}, \
  and\ \bibinfo {author} {\bibfnamefont {C.~H.}\ \bibnamefont {Lee}},\
  }\bibfield  {title} {\enquote {\bibinfo {title} {Exceptional points in
  non-hermitian systems: Applications and recent developments},}\ }\href
  {\doibase 10.1063/5.0183826} {\bibfield  {journal} {\bibinfo  {journal}
  {Appl. Phys. Lett.}\ }\textbf {\bibinfo {volume} {124}},\ \bibinfo {pages}
  {060502} (\bibinfo {year} {2024})}\BibitemShut {NoStop}%
\bibitem [{\citenamefont {El-Ganainy}\ \emph {et~al.}(2019)\citenamefont
  {El-Ganainy}, \citenamefont {Khajavikhan}, \citenamefont {Christodoulides},\
  and\ \citenamefont {Ozdemir}}]{reg2019cp}%
  \BibitemOpen
  \bibfield  {author} {\bibinfo {author} {\bibfnamefont {R.}~\bibnamefont
  {El-Ganainy}}, \bibinfo {author} {\bibfnamefont {M.}~\bibnamefont
  {Khajavikhan}}, \bibinfo {author} {\bibfnamefont {D.~N.}\ \bibnamefont
  {Christodoulides}}, \ and\ \bibinfo {author} {\bibfnamefont {S.~K.}\
  \bibnamefont {Ozdemir}},\ }\bibfield  {title} {\enquote {\bibinfo {title}
  {The dawn of non-hermitian optics},}\ }\href {\doibase
  10.1038/s42005-019-0130-z} {\bibfield  {journal} {\bibinfo  {journal}
  {Commun. Phys.}\ }\textbf {\bibinfo {volume} {2}},\ \bibinfo {pages} {37}
  (\bibinfo {year} {2019})}\BibitemShut {NoStop}%
\bibitem [{\citenamefont {Zhang}\ \emph {et~al.}(2016)\citenamefont {Zhang},
  \citenamefont {Zhang}, \citenamefont {Sheng}, \citenamefont {Yang},
  \citenamefont {Miri}, \citenamefont {Christodoulides}, \citenamefont {He},
  \citenamefont {Zhang},\ and\ \citenamefont {Xiao}}]{xiaomin2016prl}%
  \BibitemOpen
  \bibfield  {author} {\bibinfo {author} {\bibfnamefont {Z.}~\bibnamefont
  {Zhang}}, \bibinfo {author} {\bibfnamefont {Y.}~\bibnamefont {Zhang}},
  \bibinfo {author} {\bibfnamefont {J.}~\bibnamefont {Sheng}}, \bibinfo
  {author} {\bibfnamefont {L.}~\bibnamefont {Yang}}, \bibinfo {author}
  {\bibfnamefont {M.-A.}\ \bibnamefont {Miri}}, \bibinfo {author}
  {\bibfnamefont {D.~N.}\ \bibnamefont {Christodoulides}}, \bibinfo {author}
  {\bibfnamefont {B.}~\bibnamefont {He}}, \bibinfo {author} {\bibfnamefont
  {Y.}~\bibnamefont {Zhang}}, \ and\ \bibinfo {author} {\bibfnamefont
  {M.}~\bibnamefont {Xiao}},\ }\bibfield  {title} {\enquote {\bibinfo {title}
  {Observation of parity-time symmetry in optically induced atomic lattices},}\
  }\href {\doibase 10.1103/PhysRevLett.117.123601} {\bibfield  {journal}
  {\bibinfo  {journal} {Phys. Rev. Lett.}\ }\textbf {\bibinfo {volume} {117}},\
  \bibinfo {pages} {123601} (\bibinfo {year} {2016})}\BibitemShut {NoStop}%
\bibitem [{\citenamefont {Lodahl}\ \emph {et~al.}(2017)\citenamefont {Lodahl},
  \citenamefont {Mahmoodian}, \citenamefont {Stobbe}, \citenamefont
  {Rauschenbeutel}, \citenamefont {Schneeweiss}, \citenamefont {Volz},
  \citenamefont {Pichler},\ and\ \citenamefont {Zoller}}]{pzoller2017nature}%
  \BibitemOpen
  \bibfield  {author} {\bibinfo {author} {\bibfnamefont {P.}~\bibnamefont
  {Lodahl}}, \bibinfo {author} {\bibfnamefont {S.}~\bibnamefont {Mahmoodian}},
  \bibinfo {author} {\bibfnamefont {S.}~\bibnamefont {Stobbe}}, \bibinfo
  {author} {\bibfnamefont {A.}~\bibnamefont {Rauschenbeutel}}, \bibinfo
  {author} {\bibfnamefont {P.}~\bibnamefont {Schneeweiss}}, \bibinfo {author}
  {\bibfnamefont {J.}~\bibnamefont {Volz}}, \bibinfo {author} {\bibfnamefont
  {H.}~\bibnamefont {Pichler}}, \ and\ \bibinfo {author} {\bibfnamefont
  {P.}~\bibnamefont {Zoller}},\ }\bibfield  {title} {\enquote {\bibinfo {title}
  {Chiral quantum optics},}\ }\href {\doibase 10.1038/nature21037} {\bibfield
  {journal} {\bibinfo  {journal} {Nature}\ }\textbf {\bibinfo {volume} {541}},\
  \bibinfo {pages} {473} (\bibinfo {year} {2017})}\BibitemShut {NoStop}%
\bibitem [{\citenamefont {Bernier}\ \emph {et~al.}(2017)\citenamefont
  {Bernier}, \citenamefont {T\'{o}th}, \citenamefont {Koottandavida},
  \citenamefont {Ioannou}, \citenamefont {Malz}, \citenamefont {Nunnenkamp},
  \citenamefont {Feofanov},\ and\ \citenamefont {Kippenberg}}]{tjk2017nc}%
  \BibitemOpen
  \bibfield  {author} {\bibinfo {author} {\bibfnamefont {N.~R.}\ \bibnamefont
  {Bernier}}, \bibinfo {author} {\bibfnamefont {L.~D.}\ \bibnamefont
  {T\'{o}th}}, \bibinfo {author} {\bibfnamefont {A.}~\bibnamefont
  {Koottandavida}}, \bibinfo {author} {\bibfnamefont {M.~A.}\ \bibnamefont
  {Ioannou}}, \bibinfo {author} {\bibfnamefont {D.}~\bibnamefont {Malz}},
  \bibinfo {author} {\bibfnamefont {A.}~\bibnamefont {Nunnenkamp}}, \bibinfo
  {author} {\bibfnamefont {A.~K.}\ \bibnamefont {Feofanov}}, \ and\ \bibinfo
  {author} {\bibfnamefont {T.~J.}\ \bibnamefont {Kippenberg}},\ }\bibfield
  {title} {\enquote {\bibinfo {title} {Nonreciprocal reconfigurable microwave
  optomechanical circuit},}\ }\href {\doibase 10.1038/s41467-017-00447-1}
  {\bibfield  {journal} {\bibinfo  {journal} {Nat. Commun.}\ }\textbf {\bibinfo
  {volume} {8}},\ \bibinfo {pages} {604} (\bibinfo {year} {2017})}\BibitemShut
  {NoStop}%
\bibitem [{\citenamefont {Ashida}\ \emph
  {et~al.}(2020{\natexlab{b}})\citenamefont {Ashida}, \citenamefont {Gong},\
  and\ \citenamefont {Ueda}}]{mueda2020aip}%
  \BibitemOpen
  \bibfield  {author} {\bibinfo {author} {\bibfnamefont {Y.}~\bibnamefont
  {Ashida}}, \bibinfo {author} {\bibfnamefont {Z.}~\bibnamefont {Gong}}, \ and\
  \bibinfo {author} {\bibfnamefont {M.}~\bibnamefont {Ueda}},\ }\bibfield
  {title} {\enquote {\bibinfo {title} {Non-hermitian physics},}\ }\href
  {\doibase 10.1080/00018732.2021.1876991} {\bibfield  {journal} {\bibinfo
  {journal} {Adv. Phys.}\ }\textbf {\bibinfo {volume} {69}},\ \bibinfo {pages}
  {249} (\bibinfo {year} {2020}{\natexlab{b}})}\BibitemShut {NoStop}%
\bibitem [{\citenamefont {Sun}\ \emph {et~al.}(2023)\citenamefont {Sun},
  \citenamefont {Shi}, \citenamefont {Liu}, \citenamefont {Zhang},
  \citenamefont {Xiao}, \citenamefont {Jia},\ and\ \citenamefont
  {Hu}}]{yhu2023prx}%
  \BibitemOpen
  \bibfield  {author} {\bibinfo {author} {\bibfnamefont {Y.}~\bibnamefont
  {Sun}}, \bibinfo {author} {\bibfnamefont {T.}~\bibnamefont {Shi}}, \bibinfo
  {author} {\bibfnamefont {Z.}~\bibnamefont {Liu}}, \bibinfo {author}
  {\bibfnamefont {Z.}~\bibnamefont {Zhang}}, \bibinfo {author} {\bibfnamefont
  {L.}~\bibnamefont {Xiao}}, \bibinfo {author} {\bibfnamefont {S.}~\bibnamefont
  {Jia}}, \ and\ \bibinfo {author} {\bibfnamefont {Y.}~\bibnamefont {Hu}},\
  }\bibfield  {title} {\enquote {\bibinfo {title} {Fractional quantum zeno
  effect emerging from non-hermitian physics},}\ }\href {\doibase
  10.1103/PhysRevX.13.031009} {\bibfield  {journal} {\bibinfo  {journal} {Phys.
  Rev. X}\ }\textbf {\bibinfo {volume} {13}},\ \bibinfo {pages} {031009}
  (\bibinfo {year} {2023})}\BibitemShut {NoStop}%
\bibitem [{\citenamefont {Yu}\ and\ \citenamefont {Vollmer}(2021)}]{yu2021cp}%
  \BibitemOpen
  \bibfield  {author} {\bibinfo {author} {\bibfnamefont {D.}~\bibnamefont
  {Yu}}\ and\ \bibinfo {author} {\bibfnamefont {F.}~\bibnamefont {Vollmer}},\
  }\bibfield  {title} {\enquote {\bibinfo {title} {Spontaneous {PT}-symmetry
  breaking in lasing dynamics},}\ }\href {\doibase 10.1038/s42005-021-00575-7}
  {\bibfield  {journal} {\bibinfo  {journal} {Commun. Phys.}\ }\textbf
  {\bibinfo {volume} {4}},\ \bibinfo {pages} {77} (\bibinfo {year}
  {2021})}\BibitemShut {NoStop}%
\bibitem [{\citenamefont {Geva}\ \emph {et~al.}(2022)\citenamefont {Geva},
  \citenamefont {Sagie}, \citenamefont {Gershenzon}, \citenamefont {Friesem},
  \citenamefont {Davidson},\ and\ \citenamefont {Raz}}]{oraz2022sa}%
  \BibitemOpen
  \bibfield  {author} {\bibinfo {author} {\bibfnamefont {A.}~\bibnamefont
  {Geva}}, \bibinfo {author} {\bibfnamefont {G.}~\bibnamefont {Sagie}},
  \bibinfo {author} {\bibfnamefont {I.}~\bibnamefont {Gershenzon}}, \bibinfo
  {author} {\bibfnamefont {A.}~\bibnamefont {Friesem}}, \bibinfo {author}
  {\bibfnamefont {N.}~\bibnamefont {Davidson}}, \ and\ \bibinfo {author}
  {\bibfnamefont {O.}~\bibnamefont {Raz}},\ }\bibfield  {title} {\enquote
  {\bibinfo {title} {Anyonic-parity-time symmetry in complex-coupled lasers},}\
  }\href {\doibase 10.1126/sciadv.abm7454} {\bibfield  {journal} {\bibinfo
  {journal} {Sci. Adv.}\ }\textbf {\bibinfo {volume} {8}},\ \bibinfo {pages}
  {eabm7454} (\bibinfo {year} {2022})}\BibitemShut {NoStop}%
\bibitem [{\citenamefont {Zhang}\ \emph
  {et~al.}(2022{\natexlab{a}})\citenamefont {Zhang}, \citenamefont {Yang},
  \citenamefont {Sheng},\ and\ \citenamefont {Wu}}]{hbwu2022pnas}%
  \BibitemOpen
  \bibfield  {author} {\bibinfo {author} {\bibfnamefont {Q.}~\bibnamefont
  {Zhang}}, \bibinfo {author} {\bibfnamefont {C.}~\bibnamefont {Yang}},
  \bibinfo {author} {\bibfnamefont {J.}~\bibnamefont {Sheng}}, \ and\ \bibinfo
  {author} {\bibfnamefont {H.}~\bibnamefont {Wu}},\ }\bibfield  {title}
  {\enquote {\bibinfo {title} {Dissipative coupling-induced phonon lasing},}\
  }\href {\doibase 10.1073/pnas.2207543119} {\bibfield  {journal} {\bibinfo
  {journal} {PNAS}\ }\textbf {\bibinfo {volume} {119}},\ \bibinfo {pages}
  {e2207543119} (\bibinfo {year} {2022}{\natexlab{a}})}\BibitemShut {NoStop}%
\bibitem [{\citenamefont {Liang}\ \emph {et~al.}(2022)\citenamefont {Liang},
  \citenamefont {Xie}, \citenamefont {Dong}, \citenamefont {Li}, \citenamefont
  {Li}, \citenamefont {Gadway}, \citenamefont {Yi},\ and\ \citenamefont
  {Yan}}]{boyan2022prl}%
  \BibitemOpen
  \bibfield  {author} {\bibinfo {author} {\bibfnamefont {Q.}~\bibnamefont
  {Liang}}, \bibinfo {author} {\bibfnamefont {D.}~\bibnamefont {Xie}}, \bibinfo
  {author} {\bibfnamefont {Z.}~\bibnamefont {Dong}}, \bibinfo {author}
  {\bibfnamefont {H.}~\bibnamefont {Li}}, \bibinfo {author} {\bibfnamefont
  {H.}~\bibnamefont {Li}}, \bibinfo {author} {\bibfnamefont {B.}~\bibnamefont
  {Gadway}}, \bibinfo {author} {\bibfnamefont {W.}~\bibnamefont {Yi}}, \ and\
  \bibinfo {author} {\bibfnamefont {B.}~\bibnamefont {Yan}},\ }\bibfield
  {title} {\enquote {\bibinfo {title} {Dynamic signatures of non-hermitian skin
  effect and topology in ultracold atoms},}\ }\href {\doibase
  10.1103/PhysRevLett.129.070401} {\bibfield  {journal} {\bibinfo  {journal}
  {Phys. Rev. Lett.}\ }\textbf {\bibinfo {volume} {129}},\ \bibinfo {pages}
  {070401} (\bibinfo {year} {2022})}\BibitemShut {NoStop}%
\bibitem [{\citenamefont {Zhang}\ \emph
  {et~al.}(2022{\natexlab{b}})\citenamefont {Zhang}, \citenamefont {Zhang},
  \citenamefont {Lu},\ and\ \citenamefont {Chen}}]{yfchen2022aipx}%
  \BibitemOpen
  \bibfield  {author} {\bibinfo {author} {\bibfnamefont {X.}~\bibnamefont
  {Zhang}}, \bibinfo {author} {\bibfnamefont {T.}~\bibnamefont {Zhang}},
  \bibinfo {author} {\bibfnamefont {M.-H.}\ \bibnamefont {Lu}}, \ and\ \bibinfo
  {author} {\bibfnamefont {Y.-F.}\ \bibnamefont {Chen}},\ }\bibfield  {title}
  {\enquote {\bibinfo {title} {A review on non-hermitian skin effect},}\ }\href
  {\doibase 10.1080/23746149.2022.2109431} {\bibfield  {journal} {\bibinfo
  {journal} {Adv. Phys.-X}\ }\textbf {\bibinfo {volume} {7}},\ \bibinfo {pages}
  {2109431} (\bibinfo {year} {2022}{\natexlab{b}})}\BibitemShut {NoStop}%
\bibitem [{\citenamefont {Li}\ \emph {et~al.}(2024)\citenamefont {Li},
  \citenamefont {Wang}, \citenamefont {Wang}, \citenamefont {Lin},
  \citenamefont {Ma},\ and\ \citenamefont {Jiang}}]{jhjiang2024nc}%
  \BibitemOpen
  \bibfield  {author} {\bibinfo {author} {\bibfnamefont {Z.}~\bibnamefont
  {Li}}, \bibinfo {author} {\bibfnamefont {L.-W.}\ \bibnamefont {Wang}},
  \bibinfo {author} {\bibfnamefont {X.}~\bibnamefont {Wang}}, \bibinfo {author}
  {\bibfnamefont {Z.-K.}\ \bibnamefont {Lin}}, \bibinfo {author} {\bibfnamefont
  {G.}~\bibnamefont {Ma}}, \ and\ \bibinfo {author} {\bibfnamefont {J.-H.}\
  \bibnamefont {Jiang}},\ }\bibfield  {title} {\enquote {\bibinfo {title}
  {Observation of dynamic non-hermitian skin effects},}\ }\href {\doibase
  10.1038/s41467-024-50776-1} {\bibfield  {journal} {\bibinfo  {journal} {Nat.
  Commun.}\ }\textbf {\bibinfo {volume} {15}},\ \bibinfo {pages} {6544}
  (\bibinfo {year} {2024})}\BibitemShut {NoStop}%
\bibitem [{\citenamefont {Budich}\ and\ \citenamefont
  {Bergholtz}(2020)}]{jcb2020prl}%
  \BibitemOpen
  \bibfield  {author} {\bibinfo {author} {\bibfnamefont {J.~C.}\ \bibnamefont
  {Budich}}\ and\ \bibinfo {author} {\bibfnamefont {E.~J.}\ \bibnamefont
  {Bergholtz}},\ }\bibfield  {title} {\enquote {\bibinfo {title} {Non-hermitian
  topological sensors},}\ }\href {\doibase 10.1103/PhysRevLett.125.180403}
  {\bibfield  {journal} {\bibinfo  {journal} {Phys. Rev. Lett.}\ }\textbf
  {\bibinfo {volume} {125}},\ \bibinfo {pages} {180403} (\bibinfo {year}
  {2020})}\BibitemShut {NoStop}%
\bibitem [{\citenamefont {Franke}\ \emph {et~al.}(2023)\citenamefont {Franke},
  \citenamefont {Muleady}, \citenamefont {Kaubruegger}, \citenamefont {Kranzl},
  \citenamefont {Blatt}, \citenamefont {Rey}, \citenamefont {Joshi},\ and\
  \citenamefont {Roos}}]{jf2023nature}%
  \BibitemOpen
  \bibfield  {author} {\bibinfo {author} {\bibfnamefont {J.}~\bibnamefont
  {Franke}}, \bibinfo {author} {\bibfnamefont {S.~R.}\ \bibnamefont {Muleady}},
  \bibinfo {author} {\bibfnamefont {R.}~\bibnamefont {Kaubruegger}}, \bibinfo
  {author} {\bibfnamefont {F.}~\bibnamefont {Kranzl}}, \bibinfo {author}
  {\bibfnamefont {R.}~\bibnamefont {Blatt}}, \bibinfo {author} {\bibfnamefont
  {A.~M.}\ \bibnamefont {Rey}}, \bibinfo {author} {\bibfnamefont {M.~K.}\
  \bibnamefont {Joshi}}, \ and\ \bibinfo {author} {\bibfnamefont {C.~F.}\
  \bibnamefont {Roos}},\ }\bibfield  {title} {\enquote {\bibinfo {title}
  {Quantum-enhanced sensing on optical transitions through finite-range
  interactions},}\ }\href {\doibase 10.1038/s41586-023-06472-z} {\bibfield
  {journal} {\bibinfo  {journal} {Nature}\ }\textbf {\bibinfo {volume} {621}},\
  \bibinfo {pages} {740} (\bibinfo {year} {2023})}\BibitemShut {NoStop}%
\bibitem [{\citenamefont {Wiersig}(2020)}]{Wiersig2020pr}%
  \BibitemOpen
  \bibfield  {author} {\bibinfo {author} {\bibfnamefont {J.}~\bibnamefont
  {Wiersig}},\ }\bibfield  {title} {\enquote {\bibinfo {title} {Review of
  exceptional point-based sensors},}\ }\href {\doibase 10.1364/PRJ.396115}
  {\bibfield  {journal} {\bibinfo  {journal} {Photon. Res.}\ }\textbf {\bibinfo
  {volume} {8}},\ \bibinfo {pages} {1457} (\bibinfo {year} {2020})}\BibitemShut
  {NoStop}%
\bibitem [{\citenamefont {Hanai}\ \emph {et~al.}(2019)\citenamefont {Hanai},
  \citenamefont {Edelman}, \citenamefont {Ohashi},\ and\ \citenamefont
  {Littlewood}}]{pbl2019prl}%
  \BibitemOpen
  \bibfield  {author} {\bibinfo {author} {\bibfnamefont {R.}~\bibnamefont
  {Hanai}}, \bibinfo {author} {\bibfnamefont {A.}~\bibnamefont {Edelman}},
  \bibinfo {author} {\bibfnamefont {Y.}~\bibnamefont {Ohashi}}, \ and\ \bibinfo
  {author} {\bibfnamefont {P.~B.}\ \bibnamefont {Littlewood}},\ }\bibfield
  {title} {\enquote {\bibinfo {title} {Non-hermitian phase transition from a
  polariton bose-einstein condensate to a photon laser},}\ }\href {\doibase
  10.1103/PhysRevLett.122.185301} {\bibfield  {journal} {\bibinfo  {journal}
  {Phys. Rev. Lett.}\ }\textbf {\bibinfo {volume} {122}},\ \bibinfo {pages}
  {185301} (\bibinfo {year} {2019})}\BibitemShut {NoStop}%
\bibitem [{\citenamefont {Qian}\ \emph {et~al.}(2024)\citenamefont {Qian},
  \citenamefont {Li}, \citenamefont {Zhu}, \citenamefont {You},\ and\
  \citenamefont {Wang}}]{ypwang2024prl}%
  \BibitemOpen
  \bibfield  {author} {\bibinfo {author} {\bibfnamefont {J.}~\bibnamefont
  {Qian}}, \bibinfo {author} {\bibfnamefont {J.}~\bibnamefont {Li}}, \bibinfo
  {author} {\bibfnamefont {S.-Y.}\ \bibnamefont {Zhu}}, \bibinfo {author}
  {\bibfnamefont {J.~Q.}\ \bibnamefont {You}}, \ and\ \bibinfo {author}
  {\bibfnamefont {Y.-P.}\ \bibnamefont {Wang}},\ }\bibfield  {title} {\enquote
  {\bibinfo {title} {Probing $pt$-symmetry breaking of non-hermitian
  topological photonic states via strong photon-magnon coupling},}\ }\href
  {\doibase 10.1103/PhysRevLett.132.156901} {\bibfield  {journal} {\bibinfo
  {journal} {Phys. Rev. Lett.}\ }\textbf {\bibinfo {volume} {132}},\ \bibinfo
  {pages} {156901} (\bibinfo {year} {2024})}\BibitemShut {NoStop}%
\bibitem [{\citenamefont {Roccati}\ \emph {et~al.}(2024)\citenamefont
  {Roccati}, \citenamefont {Bello}, \citenamefont {Gong}, \citenamefont {Ueda},
  \citenamefont {Ciccarello}, \citenamefont {Chenu},\ and\ \citenamefont
  {Carollo}}]{ac2024nc}%
  \BibitemOpen
  \bibfield  {author} {\bibinfo {author} {\bibfnamefont {F.}~\bibnamefont
  {Roccati}}, \bibinfo {author} {\bibfnamefont {M.}~\bibnamefont {Bello}},
  \bibinfo {author} {\bibfnamefont {Z.}~\bibnamefont {Gong}}, \bibinfo {author}
  {\bibfnamefont {M.}~\bibnamefont {Ueda}}, \bibinfo {author} {\bibfnamefont
  {F.}~\bibnamefont {Ciccarello}}, \bibinfo {author} {\bibfnamefont
  {A.}~\bibnamefont {Chenu}}, \ and\ \bibinfo {author} {\bibfnamefont
  {A.}~\bibnamefont {Carollo}},\ }\bibfield  {title} {\enquote {\bibinfo
  {title} {Hermitian and non-hermitian topology from photon-mediated
  interactions},}\ }\href {\doibase 10.1038/s41467-024-46471-w} {\bibfield
  {journal} {\bibinfo  {journal} {Nat. Commun.}\ }\textbf {\bibinfo {volume}
  {15}},\ \bibinfo {pages} {2400} (\bibinfo {year} {2024})}\BibitemShut
  {NoStop}%
\bibitem [{\citenamefont {Zhang}\ \emph
  {et~al.}(2022{\natexlab{c}})\citenamefont {Zhang}, \citenamefont {Zhang},
  \citenamefont {Ding}, \citenamefont {Li}, \citenamefont {Bu}, \citenamefont
  {Wang}, \citenamefont {Yan}, \citenamefont {Su}, \citenamefont {Chen},
  \citenamefont {Nori}, \citenamefont {\"{O}zdemir}, \citenamefont {Zhou},
  \citenamefont {Jing},\ and\ \citenamefont {Feng}}]{mfeng2022nc}%
  \BibitemOpen
  \bibfield  {author} {\bibinfo {author} {\bibfnamefont {J.-W.}\ \bibnamefont
  {Zhang}}, \bibinfo {author} {\bibfnamefont {J.-Q.}\ \bibnamefont {Zhang}},
  \bibinfo {author} {\bibfnamefont {G.-Y.}\ \bibnamefont {Ding}}, \bibinfo
  {author} {\bibfnamefont {J.-C.}\ \bibnamefont {Li}}, \bibinfo {author}
  {\bibfnamefont {J.-T.}\ \bibnamefont {Bu}}, \bibinfo {author} {\bibfnamefont
  {B.}~\bibnamefont {Wang}}, \bibinfo {author} {\bibfnamefont {L.-L.}\
  \bibnamefont {Yan}}, \bibinfo {author} {\bibfnamefont {S.-L.}\ \bibnamefont
  {Su}}, \bibinfo {author} {\bibfnamefont {L.}~\bibnamefont {Chen}}, \bibinfo
  {author} {\bibfnamefont {F.}~\bibnamefont {Nori}}, \bibinfo {author}
  {\bibfnamefont {S.~K.}\ \bibnamefont {\"{O}zdemir}}, \bibinfo {author}
  {\bibfnamefont {F.}~\bibnamefont {Zhou}}, \bibinfo {author} {\bibfnamefont
  {H.}~\bibnamefont {Jing}}, \ and\ \bibinfo {author} {\bibfnamefont
  {M.}~\bibnamefont {Feng}},\ }\bibfield  {title} {\enquote {\bibinfo {title}
  {Dynamical control of quantum heat engines using exceptional points},}\
  }\href {\doibase 10.1038/s41467-022-33667-1} {\bibfield  {journal} {\bibinfo
  {journal} {Nat. Commun.}\ }\textbf {\bibinfo {volume} {13}},\ \bibinfo
  {pages} {6225} (\bibinfo {year} {2022}{\natexlab{c}})}\BibitemShut {NoStop}%
\bibitem [{\citenamefont {Bu}\ \emph {et~al.}(2023)\citenamefont {Bu},
  \citenamefont {Zhang}, \citenamefont {Ding}, \citenamefont {Li},
  \citenamefont {Zhang}, \citenamefont {Wang}, \citenamefont {Ding},
  \citenamefont {Yuan}, \citenamefont {Chen}, \citenamefont {\"{O}zdemir},
  \citenamefont {Zhou}, \citenamefont {Jing},\ and\ \citenamefont
  {Feng}}]{mfeng2023prl}%
  \BibitemOpen
  \bibfield  {author} {\bibinfo {author} {\bibfnamefont {J.-T.}\ \bibnamefont
  {Bu}}, \bibinfo {author} {\bibfnamefont {J.-Q.}\ \bibnamefont {Zhang}},
  \bibinfo {author} {\bibfnamefont {G.-Y.}\ \bibnamefont {Ding}}, \bibinfo
  {author} {\bibfnamefont {J.-C.}\ \bibnamefont {Li}}, \bibinfo {author}
  {\bibfnamefont {J.-W.}\ \bibnamefont {Zhang}}, \bibinfo {author}
  {\bibfnamefont {B.}~\bibnamefont {Wang}}, \bibinfo {author} {\bibfnamefont
  {W.-Q.}\ \bibnamefont {Ding}}, \bibinfo {author} {\bibfnamefont {W.-F.}\
  \bibnamefont {Yuan}}, \bibinfo {author} {\bibfnamefont {L.}~\bibnamefont
  {Chen}}, \bibinfo {author} {\bibfnamefont {S.~K.}\ \bibnamefont
  {\"{O}zdemir}}, \bibinfo {author} {\bibfnamefont {F.}~\bibnamefont {Zhou}},
  \bibinfo {author} {\bibfnamefont {H.}~\bibnamefont {Jing}}, \ and\ \bibinfo
  {author} {\bibfnamefont {M.}~\bibnamefont {Feng}},\ }\bibfield  {title}
  {\enquote {\bibinfo {title} {Enhancement of quantum heat engine by encircling
  a liouvillian exceptional point},}\ }\href {\doibase
  10.1103/PhysRevLett.130.110402} {\bibfield  {journal} {\bibinfo  {journal}
  {Phys. Rev. Lett.}\ }\textbf {\bibinfo {volume} {130}},\ \bibinfo {pages}
  {110402} (\bibinfo {year} {2023})}\BibitemShut {NoStop}%
\bibitem [{\citenamefont {Minganti}\ \emph {et~al.}(2019)\citenamefont
  {Minganti}, \citenamefont {Miranowicz}, \citenamefont {Chhajlany},\ and\
  \citenamefont {Nori}}]{fnori2019pra}%
  \BibitemOpen
  \bibfield  {author} {\bibinfo {author} {\bibfnamefont {F.}~\bibnamefont
  {Minganti}}, \bibinfo {author} {\bibfnamefont {A.}~\bibnamefont
  {Miranowicz}}, \bibinfo {author} {\bibfnamefont {R.~W.}\ \bibnamefont
  {Chhajlany}}, \ and\ \bibinfo {author} {\bibfnamefont {F.}~\bibnamefont
  {Nori}},\ }\bibfield  {title} {\enquote {\bibinfo {title} {Quantum
  exceptional points of non-hermitian hamiltonians and liouvillians: The
  effects of quantum jumps},}\ }\href {\doibase 10.1103/PhysRevA.100.062131}
  {\bibfield  {journal} {\bibinfo  {journal} {Phys. Rev. A}\ }\textbf {\bibinfo
  {volume} {100}},\ \bibinfo {pages} {062131} (\bibinfo {year}
  {2019})}\BibitemShut {NoStop}%
\bibitem [{\citenamefont {Minganti}\ \emph {et~al.}(2018)\citenamefont
  {Minganti}, \citenamefont {Biella}, \citenamefont {Bartolo},\ and\
  \citenamefont {Ciuti}}]{cc2018pra}%
  \BibitemOpen
  \bibfield  {author} {\bibinfo {author} {\bibfnamefont {F.}~\bibnamefont
  {Minganti}}, \bibinfo {author} {\bibfnamefont {A.}~\bibnamefont {Biella}},
  \bibinfo {author} {\bibfnamefont {N.}~\bibnamefont {Bartolo}}, \ and\
  \bibinfo {author} {\bibfnamefont {C.}~\bibnamefont {Ciuti}},\ }\bibfield
  {title} {\enquote {\bibinfo {title} {Spectral theory of liouvillians for
  dissipative phase transitions},}\ }\href {\doibase
  10.1103/PhysRevA.98.042118} {\bibfield  {journal} {\bibinfo  {journal} {Phys.
  Rev. A}\ }\textbf {\bibinfo {volume} {98}},\ \bibinfo {pages} {042118}
  (\bibinfo {year} {2018})}\BibitemShut {NoStop}%
\bibitem [{\citenamefont {Quijandr\'{i}a}\ \emph {et~al.}(2018)\citenamefont
  {Quijandr\'{i}a}, \citenamefont {Naether}, \citenamefont {\"{O}zdemir},
  \citenamefont {Nori},\ and\ \citenamefont {Zueco}}]{dzueco2018pra}%
  \BibitemOpen
  \bibfield  {author} {\bibinfo {author} {\bibfnamefont {F.}~\bibnamefont
  {Quijandr\'{i}a}}, \bibinfo {author} {\bibfnamefont {U.}~\bibnamefont
  {Naether}}, \bibinfo {author} {\bibfnamefont {S.~K.}\ \bibnamefont
  {\"{O}zdemir}}, \bibinfo {author} {\bibfnamefont {F.}~\bibnamefont {Nori}}, \
  and\ \bibinfo {author} {\bibfnamefont {D.}~\bibnamefont {Zueco}},\ }\bibfield
   {title} {\enquote {\bibinfo {title} {$\mathcal{PT}$-symmetric circuit
  qed},}\ }\href {\doibase 10.1103/PhysRevA.97.053846} {\bibfield  {journal}
  {\bibinfo  {journal} {Phys. Rev. A}\ }\textbf {\bibinfo {volume} {97}},\
  \bibinfo {pages} {053846} (\bibinfo {year} {2018})}\BibitemShut {NoStop}%
\bibitem [{\citenamefont {Dicke}(1954)}]{dicke1954pr}%
  \BibitemOpen
  \bibfield  {author} {\bibinfo {author} {\bibfnamefont {R.~H.}\ \bibnamefont
  {Dicke}},\ }\bibfield  {title} {\enquote {\bibinfo {title} {Coherence in
  spontaneous radiation processes},}\ }\href {\doibase 10.1103/PhysRev.93.99}
  {\bibfield  {journal} {\bibinfo  {journal} {Phys. Rev.}\ }\textbf {\bibinfo
  {volume} {93}},\ \bibinfo {pages} {99} (\bibinfo {year} {1954})}\BibitemShut
  {NoStop}%
\bibitem [{\citenamefont {Kirton}\ \emph {et~al.}(2019)\citenamefont {Kirton},
  \citenamefont {Roses}, \citenamefont {Keeling},\ and\ \citenamefont
  {Dalla~Torre}}]{pkirton2019aqt}%
  \BibitemOpen
  \bibfield  {author} {\bibinfo {author} {\bibfnamefont {P.}~\bibnamefont
  {Kirton}}, \bibinfo {author} {\bibfnamefont {M.~M.}\ \bibnamefont {Roses}},
  \bibinfo {author} {\bibfnamefont {J.}~\bibnamefont {Keeling}}, \ and\
  \bibinfo {author} {\bibfnamefont {E.~G.}\ \bibnamefont {Dalla~Torre}},\
  }\bibfield  {title} {\enquote {\bibinfo {title} {Introduction to the dicke
  model: From equilibrium to nonequilibrium, and vice versa},}\ }\href
  {\doibase https://doi.org/10.1002/qute.201800043} {\bibfield  {journal}
  {\bibinfo  {journal} {Adv. Quantum Technol.}\ }\textbf {\bibinfo {volume}
  {2}},\ \bibinfo {pages} {1800043} (\bibinfo {year} {2019})}\BibitemShut
  {NoStop}%
\bibitem [{\citenamefont {Chu}\ \emph {et~al.}(2021)\citenamefont {Chu},
  \citenamefont {Zhang}, \citenamefont {Yu},\ and\ \citenamefont
  {Cai}}]{jmcai2021prl}%
  \BibitemOpen
  \bibfield  {author} {\bibinfo {author} {\bibfnamefont {Y.~M.}\ \bibnamefont
  {Chu}}, \bibinfo {author} {\bibfnamefont {S.~L.}\ \bibnamefont {Zhang}},
  \bibinfo {author} {\bibfnamefont {B.~Y.}\ \bibnamefont {Yu}}, \ and\ \bibinfo
  {author} {\bibfnamefont {J.~M.}\ \bibnamefont {Cai}},\ }\bibfield  {title}
  {\enquote {\bibinfo {title} {Dynamic framework for criticality-enhanced
  quantum sensing},}\ }\href {\doibase 10.1103/PhysRevLett.126.010502}
  {\bibfield  {journal} {\bibinfo  {journal} {Phys. Rev. Lett.}\ }\textbf
  {\bibinfo {volume} {126}},\ \bibinfo {pages} {010502} (\bibinfo {year}
  {2021})}\BibitemShut {NoStop}%
\bibitem [{\citenamefont {Ilias}\ \emph {et~al.}(2022)\citenamefont {Ilias},
  \citenamefont {Yang}, \citenamefont {Huelga},\ and\ \citenamefont
  {Plenio}}]{mplenio2022prx}%
  \BibitemOpen
  \bibfield  {author} {\bibinfo {author} {\bibfnamefont {T.}~\bibnamefont
  {Ilias}}, \bibinfo {author} {\bibfnamefont {D.~Y.}\ \bibnamefont {Yang}},
  \bibinfo {author} {\bibfnamefont {S.~F.}\ \bibnamefont {Huelga}}, \ and\
  \bibinfo {author} {\bibfnamefont {M.~B.}\ \bibnamefont {Plenio}},\ }\bibfield
   {title} {\enquote {\bibinfo {title} {Criticality-enhanced quantum sensing
  via continuous measurement},}\ }\href {\doibase 10.1103/PRXQuantum.3.010354}
  {\bibfield  {journal} {\bibinfo  {journal} {PRX Quantum}\ }\textbf {\bibinfo
  {volume} {3}},\ \bibinfo {pages} {010354} (\bibinfo {year}
  {2022})}\BibitemShut {NoStop}%
\bibitem [{\citenamefont {Lambert}\ \emph {et~al.}(2004)\citenamefont
  {Lambert}, \citenamefont {Emary},\ and\ \citenamefont
  {Brandes}}]{lambert2004prl}%
  \BibitemOpen
  \bibfield  {author} {\bibinfo {author} {\bibfnamefont {N.}~\bibnamefont
  {Lambert}}, \bibinfo {author} {\bibfnamefont {C.}~\bibnamefont {Emary}}, \
  and\ \bibinfo {author} {\bibfnamefont {T.}~\bibnamefont {Brandes}},\
  }\bibfield  {title} {\enquote {\bibinfo {title} {Entanglement and the phase
  transition in single-mode superradiance},}\ }\href {\doibase
  10.1103/PhysRevLett.92.073602} {\bibfield  {journal} {\bibinfo  {journal}
  {Phys. Rev. Lett.}\ }\textbf {\bibinfo {volume} {92}},\ \bibinfo {pages}
  {073602} (\bibinfo {year} {2004})}\BibitemShut {NoStop}%
\bibitem [{\citenamefont {Das}\ \emph {et~al.}(2023)\citenamefont {Das},
  \citenamefont {Bhakuni},\ and\ \citenamefont {Sharma}}]{das2023pra}%
  \BibitemOpen
  \bibfield  {author} {\bibinfo {author} {\bibfnamefont {P.}~\bibnamefont
  {Das}}, \bibinfo {author} {\bibfnamefont {D.~S.}\ \bibnamefont {Bhakuni}}, \
  and\ \bibinfo {author} {\bibfnamefont {A.}~\bibnamefont {Sharma}},\
  }\bibfield  {title} {\enquote {\bibinfo {title} {Phase transitions of the
  anisotropic dicke model},}\ }\href {\doibase 10.1103/PhysRevA.107.043706}
  {\bibfield  {journal} {\bibinfo  {journal} {Phys. Rev. A}\ }\textbf {\bibinfo
  {volume} {107}},\ \bibinfo {pages} {043706} (\bibinfo {year}
  {2023})}\BibitemShut {NoStop}%
\bibitem [{\citenamefont {Garbe}\ \emph {et~al.}(2017)\citenamefont {Garbe},
  \citenamefont {Egusquiza}, \citenamefont {Solano}, \citenamefont {Ciuti},
  \citenamefont {Coudreau}, \citenamefont {Milman},\ and\ \citenamefont
  {Felicetti}}]{garbe2017pra}%
  \BibitemOpen
  \bibfield  {author} {\bibinfo {author} {\bibfnamefont {L.}~\bibnamefont
  {Garbe}}, \bibinfo {author} {\bibfnamefont {I.~L.}\ \bibnamefont
  {Egusquiza}}, \bibinfo {author} {\bibfnamefont {E.}~\bibnamefont {Solano}},
  \bibinfo {author} {\bibfnamefont {C.}~\bibnamefont {Ciuti}}, \bibinfo
  {author} {\bibfnamefont {T.}~\bibnamefont {Coudreau}}, \bibinfo {author}
  {\bibfnamefont {P.}~\bibnamefont {Milman}}, \ and\ \bibinfo {author}
  {\bibfnamefont {S.}~\bibnamefont {Felicetti}},\ }\bibfield  {title} {\enquote
  {\bibinfo {title} {Superradiant phase transition in the ultrastrong-coupling
  regime of the two-photon dicke model},}\ }\href {\doibase
  10.1103/PhysRevA.95.053854} {\bibfield  {journal} {\bibinfo  {journal} {Phys.
  Rev. A}\ }\textbf {\bibinfo {volume} {95}},\ \bibinfo {pages} {053854}
  (\bibinfo {year} {2017})}\BibitemShut {NoStop}%
\bibitem [{\citenamefont {Deguchi}\ and\ \citenamefont
  {Ghosh}(2009)}]{kpg2009pre}%
  \BibitemOpen
  \bibfield  {author} {\bibinfo {author} {\bibfnamefont {T.}~\bibnamefont
  {Deguchi}}\ and\ \bibinfo {author} {\bibfnamefont {P.~K.}\ \bibnamefont
  {Ghosh}},\ }\bibfield  {title} {\enquote {\bibinfo {title} {Quantum phase
  transition in a pseudo-hermitian dicke model},}\ }\href {\doibase
  10.1103/PhysRevE.80.021107} {\bibfield  {journal} {\bibinfo  {journal} {Phys.
  Rev. E}\ }\textbf {\bibinfo {volume} {80}},\ \bibinfo {pages} {021107}
  (\bibinfo {year} {2009})}\BibitemShut {NoStop}%
\bibitem [{\citenamefont {Chiacchio}\ \emph {et~al.}(2023)\citenamefont
  {Chiacchio}, \citenamefont {Nunnenkamp},\ and\ \citenamefont
  {Brunelli}}]{mb2023prl}%
  \BibitemOpen
  \bibfield  {author} {\bibinfo {author} {\bibfnamefont {E.~I.~R.}\
  \bibnamefont {Chiacchio}}, \bibinfo {author} {\bibfnamefont {A.}~\bibnamefont
  {Nunnenkamp}}, \ and\ \bibinfo {author} {\bibfnamefont {M.}~\bibnamefont
  {Brunelli}},\ }\bibfield  {title} {\enquote {\bibinfo {title} {Nonreciprocal
  dicke model},}\ }\href {\doibase 10.1103/PhysRevLett.131.113602} {\bibfield
  {journal} {\bibinfo  {journal} {Phys. Rev. Lett.}\ }\textbf {\bibinfo
  {volume} {131}},\ \bibinfo {pages} {113602} (\bibinfo {year}
  {2023})}\BibitemShut {NoStop}%
\bibitem [{\citenamefont {Lu}\ \emph {et~al.}(2023)\citenamefont {Lu},
  \citenamefont {Li}, \citenamefont {Shi}, \citenamefont {Fan}, \citenamefont
  {Mangazeev}, \citenamefont {Li},\ and\ \citenamefont
  {Batchelor}}]{zmli2023pra}%
  \BibitemOpen
  \bibfield  {author} {\bibinfo {author} {\bibfnamefont {X.}~\bibnamefont
  {Lu}}, \bibinfo {author} {\bibfnamefont {H.}~\bibnamefont {Li}}, \bibinfo
  {author} {\bibfnamefont {J.-K.}\ \bibnamefont {Shi}}, \bibinfo {author}
  {\bibfnamefont {L.-B.}\ \bibnamefont {Fan}}, \bibinfo {author} {\bibfnamefont
  {V.}~\bibnamefont {Mangazeev}}, \bibinfo {author} {\bibfnamefont {Z.-M.}\
  \bibnamefont {Li}}, \ and\ \bibinfo {author} {\bibfnamefont {M.~T.}\
  \bibnamefont {Batchelor}},\ }\bibfield  {title} {\enquote {\bibinfo {title}
  {$\mathcal{PT}$-symmetric quantum rabi model},}\ }\href {\doibase
  10.1103/PhysRevA.108.053712} {\bibfield  {journal} {\bibinfo  {journal}
  {Phys. Rev. A}\ }\textbf {\bibinfo {volume} {108}},\ \bibinfo {pages}
  {053712} (\bibinfo {year} {2023})}\BibitemShut {NoStop}%
\bibitem [{\citenamefont {Mandal}\ and\ \citenamefont
  {Bergholtz}(2021)}]{HOEPtheory1}%
  \BibitemOpen
  \bibfield  {author} {\bibinfo {author} {\bibfnamefont {I.}~\bibnamefont
  {Mandal}}\ and\ \bibinfo {author} {\bibfnamefont {E.~J.}\ \bibnamefont
  {Bergholtz}},\ }\bibfield  {title} {\enquote {\bibinfo {title} {Symmetry and
  higher-order exceptional points},}\ }\href {\doibase
  10.1103/PhysRevLett.127.186601} {\bibfield  {journal} {\bibinfo  {journal}
  {Phys. Rev. Lett.}\ }\textbf {\bibinfo {volume} {127}},\ \bibinfo {pages}
  {186601} (\bibinfo {year} {2021})}\BibitemShut {NoStop}%
\bibitem [{\citenamefont {Hodaei}\ \emph {et~al.}(2017)\citenamefont {Hodaei},
  \citenamefont {Ul~Hassan}, \citenamefont {Wittek}, \citenamefont
  {Garcia-Gracia}, \citenamefont {El-Ganainy}, \citenamefont
  {Christodoulides},\ and\ \citenamefont {Khajavikhan}}]{HOEP7}%
  \BibitemOpen
  \bibfield  {author} {\bibinfo {author} {\bibfnamefont {H.}~\bibnamefont
  {Hodaei}}, \bibinfo {author} {\bibfnamefont {A.}~\bibnamefont {Ul~Hassan}},
  \bibinfo {author} {\bibfnamefont {S.}~\bibnamefont {Wittek}}, \bibinfo
  {author} {\bibfnamefont {H.}~\bibnamefont {Garcia-Gracia}}, \bibinfo {author}
  {\bibfnamefont {R.}~\bibnamefont {El-Ganainy}}, \bibinfo {author}
  {\bibfnamefont {D.~N.}\ \bibnamefont {Christodoulides}}, \ and\ \bibinfo
  {author} {\bibfnamefont {M.}~\bibnamefont {Khajavikhan}},\ }\bibfield
  {title} {\enquote {\bibinfo {title} {Enhanced sensitivity at higher-order
  exceptional points},}\ }\href {\doibase 10.1038/nature23280} {\bibfield
  {journal} {\bibinfo  {journal} {Nature}\ }\textbf {\bibinfo {volume} {548}},\
  \bibinfo {pages} {187} (\bibinfo {year} {2017})}\BibitemShut {NoStop}%
\bibitem [{\citenamefont {Fan}\ \emph {et~al.}(2024)\citenamefont {Fan},
  \citenamefont {Zhang}, \citenamefont {Shan}, \citenamefont {Yang},
  \citenamefont {Yao}, \citenamefont {Lin},\ and\ \citenamefont {Liu}}]{HOEP9}%
  \BibitemOpen
  \bibfield  {author} {\bibinfo {author} {\bibfnamefont {M.~S.}\ \bibnamefont
  {Fan}}, \bibinfo {author} {\bibfnamefont {H.}~\bibnamefont {Zhang}}, \bibinfo
  {author} {\bibfnamefont {X.~L.}\ \bibnamefont {Shan}}, \bibinfo {author}
  {\bibfnamefont {M.~L.}\ \bibnamefont {Yang}}, \bibinfo {author}
  {\bibfnamefont {Y.}~\bibnamefont {Yao}}, \bibinfo {author} {\bibfnamefont
  {W.}~\bibnamefont {Lin}}, \ and\ \bibinfo {author} {\bibfnamefont
  {B.}~\bibnamefont {Liu}},\ }\bibfield  {title} {\enquote {\bibinfo {title}
  {Sensitivity enhancement at higher order exceptional point based on coupled
  whispering gallery modes in microstructured optical fibers},}\ }\href
  {\doibase https://doi.org/10.1016/j.optlastec.2023.110434} {\bibfield
  {journal} {\bibinfo  {journal} {Opt. Laser Technol.}\ }\textbf {\bibinfo
  {volume} {172}},\ \bibinfo {pages} {110434} (\bibinfo {year}
  {2024})}\BibitemShut {NoStop}%
\bibitem [{\citenamefont {Holstein}\ and\ \citenamefont
  {Primakoff}(1940)}]{HPtransformation}%
  \BibitemOpen
  \bibfield  {author} {\bibinfo {author} {\bibfnamefont {T.}~\bibnamefont
  {Holstein}}\ and\ \bibinfo {author} {\bibfnamefont {H.}~\bibnamefont
  {Primakoff}},\ }\bibfield  {title} {\enquote {\bibinfo {title} {Field
  dependence of the intrinsic domain magnetization of a ferromagnet},}\ }\href
  {\doibase 10.1103/PhysRev.58.1098} {\bibfield  {journal} {\bibinfo  {journal}
  {Phys. Rev.}\ }\textbf {\bibinfo {volume} {58}},\ \bibinfo {pages} {1098}
  (\bibinfo {year} {1940})}\BibitemShut {NoStop}%
\bibitem [{\citenamefont {Ding}\ \emph {et~al.}(2016)\citenamefont {Ding},
  \citenamefont {Ma}, \citenamefont {Xiao}, \citenamefont {Zhang},\ and\
  \citenamefont {Chan}}]{HOEP6}%
  \BibitemOpen
  \bibfield  {author} {\bibinfo {author} {\bibfnamefont {K.}~\bibnamefont
  {Ding}}, \bibinfo {author} {\bibfnamefont {G.~C.}\ \bibnamefont {Ma}},
  \bibinfo {author} {\bibfnamefont {M.}~\bibnamefont {Xiao}}, \bibinfo {author}
  {\bibfnamefont {Z.~Q.}\ \bibnamefont {Zhang}}, \ and\ \bibinfo {author}
  {\bibfnamefont {C.~T.}\ \bibnamefont {Chan}},\ }\bibfield  {title} {\enquote
  {\bibinfo {title} {Emergence, coalescence, and topological properties of
  multiple exceptional points and their experimental realization},}\ }\href
  {\doibase 10.1103/PhysRevX.6.021007} {\bibfield  {journal} {\bibinfo
  {journal} {Phys. Rev. X}\ }\textbf {\bibinfo {volume} {6}},\ \bibinfo {pages}
  {021007} (\bibinfo {year} {2016})}\BibitemShut {NoStop}%
\bibitem [{\citenamefont {Zhang}\ and\ \citenamefont {Chan}(2018)}]{HOEP1}%
  \BibitemOpen
  \bibfield  {author} {\bibinfo {author} {\bibfnamefont {X.-L.}\ \bibnamefont
  {Zhang}}\ and\ \bibinfo {author} {\bibfnamefont {C.~T.}\ \bibnamefont
  {Chan}},\ }\bibfield  {title} {\enquote {\bibinfo {title} {Hybrid exceptional
  point and its dynamical encircling in a two-state system},}\ }\href {\doibase
  10.1103/PhysRevA.98.033810} {\bibfield  {journal} {\bibinfo  {journal} {Phys.
  Rev. A}\ }\textbf {\bibinfo {volume} {98}},\ \bibinfo {pages} {033810}
  (\bibinfo {year} {2018})}\BibitemShut {NoStop}%
\bibitem [{\citenamefont {Tang}\ \emph {et~al.}(2020)\citenamefont {Tang},
  \citenamefont {Jiang}, \citenamefont {Ding}, \citenamefont {Xiao},
  \citenamefont {Zhang}, \citenamefont {Chan},\ and\ \citenamefont
  {Ma}}]{HOEP2}%
  \BibitemOpen
  \bibfield  {author} {\bibinfo {author} {\bibfnamefont {W.}~\bibnamefont
  {Tang}}, \bibinfo {author} {\bibfnamefont {X.}~\bibnamefont {Jiang}},
  \bibinfo {author} {\bibfnamefont {K.}~\bibnamefont {Ding}}, \bibinfo {author}
  {\bibfnamefont {Y.-X.}\ \bibnamefont {Xiao}}, \bibinfo {author}
  {\bibfnamefont {Z.-Q.}\ \bibnamefont {Zhang}}, \bibinfo {author}
  {\bibfnamefont {C.~T.}\ \bibnamefont {Chan}}, \ and\ \bibinfo {author}
  {\bibfnamefont {G.~C.}\ \bibnamefont {Ma}},\ }\bibfield  {title} {\enquote
  {\bibinfo {title} {Exceptional nexus with a hybrid topological invariant},}\
  }\href {\doibase 10.1126/science.abd8872} {\bibfield  {journal} {\bibinfo
  {journal} {Science}\ }\textbf {\bibinfo {volume} {370}},\ \bibinfo {pages}
  {1077} (\bibinfo {year} {2020})}\BibitemShut {NoStop}%
\bibitem [{\citenamefont {Sayyad}\ and\ \citenamefont
  {Kunst}(2022)}]{HOEPtheory2}%
  \BibitemOpen
  \bibfield  {author} {\bibinfo {author} {\bibfnamefont {S.}~\bibnamefont
  {Sayyad}}\ and\ \bibinfo {author} {\bibfnamefont {F.~K.}\ \bibnamefont
  {Kunst}},\ }\bibfield  {title} {\enquote {\bibinfo {title} {Realizing
  exceptional points of any order in the presence of symmetry},}\ }\href
  {\doibase 10.1103/PhysRevResearch.4.023130} {\bibfield  {journal} {\bibinfo
  {journal} {Phys. Rev. Res.}\ }\textbf {\bibinfo {volume} {4}},\ \bibinfo
  {pages} {023130} (\bibinfo {year} {2022})}\BibitemShut {NoStop}%
\bibitem [{\citenamefont {Feng}\ \emph {et~al.}(2017)\citenamefont {Feng},
  \citenamefont {El-Ganainy},\ and\ \citenamefont {Ge}}]{HOEPphoton1}%
  \BibitemOpen
  \bibfield  {author} {\bibinfo {author} {\bibfnamefont {L.}~\bibnamefont
  {Feng}}, \bibinfo {author} {\bibfnamefont {R.}~\bibnamefont {El-Ganainy}}, \
  and\ \bibinfo {author} {\bibfnamefont {L.}~\bibnamefont {Ge}},\ }\bibfield
  {title} {\enquote {\bibinfo {title} {Non-hermitian photonics based on
  parity–time symmetry},}\ }\href
  {https://www.nature.com/articles/s41566-017-0031-1} {\bibfield  {journal}
  {\bibinfo  {journal} {Nat. Photon.}\ }\textbf {\bibinfo {volume} {11}},\
  \bibinfo {pages} {752} (\bibinfo {year} {2017})}\BibitemShut {NoStop}%
\bibitem [{\citenamefont {Miri}\ and\ \citenamefont
  {Al{\`u}}(2019)}]{HOEPphoton2}%
  \BibitemOpen
  \bibfield  {author} {\bibinfo {author} {\bibfnamefont {M.-A.}\ \bibnamefont
  {Miri}}\ and\ \bibinfo {author} {\bibfnamefont {A.}~\bibnamefont {Al{\`u}}},\
  }\bibfield  {title} {\enquote {\bibinfo {title} {Exceptional points in optics
  and photonics},}\ }\href
  {https://www.science.org/doi/abs/10.1126/science.aar7709} {\bibfield
  {journal} {\bibinfo  {journal} {Science}\ }\textbf {\bibinfo {volume}
  {363}},\ \bibinfo {pages} {eaar7709} (\bibinfo {year} {2019})}\BibitemShut
  {NoStop}%
\bibitem [{\citenamefont {\"{O}zdemir}\ \emph {et~al.}(2019)\citenamefont
  {\"{O}zdemir}, \citenamefont {Rotter}, \citenamefont {Nori},\ and\
  \citenamefont {Yang}}]{HOEPphoton3}%
  \BibitemOpen
  \bibfield  {author} {\bibinfo {author} {\bibfnamefont {S.~K.}\ \bibnamefont
  {\"{O}zdemir}}, \bibinfo {author} {\bibfnamefont {S.}~\bibnamefont {Rotter}},
  \bibinfo {author} {\bibfnamefont {F.}~\bibnamefont {Nori}}, \ and\ \bibinfo
  {author} {\bibfnamefont {L.}~\bibnamefont {Yang}},\ }\bibfield  {title}
  {\enquote {\bibinfo {title} {Parity-time symmetry and exceptional points in
  photonics},}\ }\href {https://www.nature.com/articles/s41563-019-0304-9}
  {\bibfield  {journal} {\bibinfo  {journal} {Nat. Mater.}\ }\textbf {\bibinfo
  {volume} {18}},\ \bibinfo {pages} {783} (\bibinfo {year} {2019})}\BibitemShut
  {NoStop}%
\bibitem [{\citenamefont {Li}\ \emph {et~al.}(2023)\citenamefont {Li},
  \citenamefont {Wei}, \citenamefont {Cotrufo}, \citenamefont {Chen},
  \citenamefont {Mann}, \citenamefont {Ni}, \citenamefont {Xu}, \citenamefont
  {Chen}, \citenamefont {Wang}, \citenamefont {Fan}, \citenamefont {Qiu},
  \citenamefont {Al{\`u}},\ and\ \citenamefont {Chen}}]{HOEPphoton4}%
  \BibitemOpen
  \bibfield  {author} {\bibinfo {author} {\bibfnamefont {A.}~\bibnamefont
  {Li}}, \bibinfo {author} {\bibfnamefont {H.}~\bibnamefont {Wei}}, \bibinfo
  {author} {\bibfnamefont {M.}~\bibnamefont {Cotrufo}}, \bibinfo {author}
  {\bibfnamefont {W.}~\bibnamefont {Chen}}, \bibinfo {author} {\bibfnamefont
  {S.~A.}\ \bibnamefont {Mann}}, \bibinfo {author} {\bibfnamefont
  {X.}~\bibnamefont {Ni}}, \bibinfo {author} {\bibfnamefont {B.}~\bibnamefont
  {Xu}}, \bibinfo {author} {\bibfnamefont {J.}~\bibnamefont {Chen}}, \bibinfo
  {author} {\bibfnamefont {J.}~\bibnamefont {Wang}}, \bibinfo {author}
  {\bibfnamefont {S.}~\bibnamefont {Fan}}, \bibinfo {author} {\bibfnamefont
  {C.-W.}\ \bibnamefont {Qiu}}, \bibinfo {author} {\bibfnamefont
  {A.}~\bibnamefont {Al{\`u}}}, \ and\ \bibinfo {author} {\bibfnamefont
  {L.}~\bibnamefont {Chen}},\ }\bibfield  {title} {\enquote {\bibinfo {title}
  {Exceptional points and non-hermitian photonics at the nanoscale},}\ }\href
  {https://www.nature.com/articles/s41565-023-01408-0} {\bibfield  {journal}
  {\bibinfo  {journal} {Nat. Nanotechnol.}\ }\textbf {\bibinfo {volume} {18}},\
  \bibinfo {pages} {706} (\bibinfo {year} {2023})}\BibitemShut {NoStop}%
\bibitem [{\citenamefont {Liang}\ \emph {et~al.}(2023)\citenamefont {Liang},
  \citenamefont {Tang}, \citenamefont {Xu},\ and\ \citenamefont
  {Liu}}]{HOEPquantum1}%
  \BibitemOpen
  \bibfield  {author} {\bibinfo {author} {\bibfnamefont {C.}~\bibnamefont
  {Liang}}, \bibinfo {author} {\bibfnamefont {Y.}~\bibnamefont {Tang}},
  \bibinfo {author} {\bibfnamefont {A.-N.}\ \bibnamefont {Xu}}, \ and\ \bibinfo
  {author} {\bibfnamefont {Y.-C.}\ \bibnamefont {Liu}},\ }\bibfield  {title}
  {\enquote {\bibinfo {title} {Observation of exceptional points in thermal
  atomic ensembles},}\ }\href {\doibase 10.1103/PhysRevLett.130.263601}
  {\bibfield  {journal} {\bibinfo  {journal} {Phys. Rev. Lett.}\ }\textbf
  {\bibinfo {volume} {130}},\ \bibinfo {pages} {263601} (\bibinfo {year}
  {2023})}\BibitemShut {NoStop}%
\bibitem [{\citenamefont {Wu}\ \emph {et~al.}(2024)\citenamefont {Wu},
  \citenamefont {Wang}, \citenamefont {Ye}, \citenamefont {Liu}, \citenamefont
  {Niu}, \citenamefont {Duan}, \citenamefont {Wang}, \citenamefont {Rong},\
  and\ \citenamefont {Du}}]{HOEPquantum2}%
  \BibitemOpen
  \bibfield  {author} {\bibinfo {author} {\bibfnamefont {Y.}~\bibnamefont
  {Wu}}, \bibinfo {author} {\bibfnamefont {Y.}~\bibnamefont {Wang}}, \bibinfo
  {author} {\bibfnamefont {X.}~\bibnamefont {Ye}}, \bibinfo {author}
  {\bibfnamefont {W.}~\bibnamefont {Liu}}, \bibinfo {author} {\bibfnamefont
  {Z.}~\bibnamefont {Niu}}, \bibinfo {author} {\bibfnamefont {C.-K.}\
  \bibnamefont {Duan}}, \bibinfo {author} {\bibfnamefont {Y.}~\bibnamefont
  {Wang}}, \bibinfo {author} {\bibfnamefont {X.}~\bibnamefont {Rong}}, \ and\
  \bibinfo {author} {\bibfnamefont {J.}~\bibnamefont {Du}},\ }\bibfield
  {title} {\enquote {\bibinfo {title} {Third-order exceptional line in a
  nitrogen-vacancy spin system},}\ }\href
  {https://www.nature.com/articles/s41565-023-01583-0} {\bibfield  {journal}
  {\bibinfo  {journal} {Nat. Nanotechnol.}\ }\textbf {\bibinfo {volume} {19}},\
  \bibinfo {pages} {160} (\bibinfo {year} {2024})}\BibitemShut {NoStop}%
\bibitem [{\citenamefont {Wang}\ \emph {et~al.}(2021)\citenamefont {Wang},
  \citenamefont {Guo},\ and\ \citenamefont {Berakdar}}]{HOEP8}%
  \BibitemOpen
  \bibfield  {author} {\bibinfo {author} {\bibfnamefont {X.~G.}\ \bibnamefont
  {Wang}}, \bibinfo {author} {\bibfnamefont {G.~H.}\ \bibnamefont {Guo}}, \
  and\ \bibinfo {author} {\bibfnamefont {J.}~\bibnamefont {Berakdar}},\
  }\bibfield  {title} {\enquote {\bibinfo {title} {Enhanced sensitivity at
  magnetic high-order exceptional points and topological energy transfer in
  magnonic planar waveguides},}\ }\href {\doibase
  10.1103/PhysRevApplied.15.034050} {\bibfield  {journal} {\bibinfo  {journal}
  {Phys. Rev. Appl.}\ }\textbf {\bibinfo {volume} {15}},\ \bibinfo {pages}
  {034050} (\bibinfo {year} {2021})}\BibitemShut {NoStop}%
\bibitem [{\citenamefont {Lin}\ \emph {et~al.}(2016)\citenamefont {Lin},
  \citenamefont {Pick}, \citenamefont {Lončar},\ and\ \citenamefont
  {Rodriguez}}]{HOEPemission1}%
  \BibitemOpen
  \bibfield  {author} {\bibinfo {author} {\bibfnamefont {Z.}~\bibnamefont
  {Lin}}, \bibinfo {author} {\bibfnamefont {A.}~\bibnamefont {Pick}}, \bibinfo
  {author} {\bibfnamefont {M.}~\bibnamefont {Lončar}}, \ and\ \bibinfo
  {author} {\bibfnamefont {A.~W.}\ \bibnamefont {Rodriguez}},\ }\bibfield
  {title} {\enquote {\bibinfo {title} {Enhanced spontaneous emission at
  third-order dirac exceptional points in inverse-designed photonic
  crystals},}\ }\href {\doibase 10.1103/PhysRevLett.117.107402} {\bibfield
  {journal} {\bibinfo  {journal} {Phys. Rev. Lett.}\ }\textbf {\bibinfo
  {volume} {117}},\ \bibinfo {pages} {107402} (\bibinfo {year}
  {2016})}\BibitemShut {NoStop}%
\bibitem [{\citenamefont {Ferrier}\ \emph {et~al.}(2022)\citenamefont
  {Ferrier}, \citenamefont {Bouteyre}, \citenamefont {Pick}, \citenamefont
  {Cueff}, \citenamefont {Dang}, \citenamefont {Diederichs}, \citenamefont
  {Belarouci}, \citenamefont {Benyattou}, \citenamefont {Zhao}, \citenamefont
  {Su}, \citenamefont {Xing}, \citenamefont {Xiong},\ and\ \citenamefont
  {Nguyen}}]{HOEPemission2}%
  \BibitemOpen
  \bibfield  {author} {\bibinfo {author} {\bibfnamefont {L.}~\bibnamefont
  {Ferrier}}, \bibinfo {author} {\bibfnamefont {P.}~\bibnamefont {Bouteyre}},
  \bibinfo {author} {\bibfnamefont {A.}~\bibnamefont {Pick}}, \bibinfo {author}
  {\bibfnamefont {S.}~\bibnamefont {Cueff}}, \bibinfo {author} {\bibfnamefont
  {N.~H.~M.}\ \bibnamefont {Dang}}, \bibinfo {author} {\bibfnamefont
  {C.}~\bibnamefont {Diederichs}}, \bibinfo {author} {\bibfnamefont
  {A.}~\bibnamefont {Belarouci}}, \bibinfo {author} {\bibfnamefont
  {T.}~\bibnamefont {Benyattou}}, \bibinfo {author} {\bibfnamefont {J.~X.}\
  \bibnamefont {Zhao}}, \bibinfo {author} {\bibfnamefont {R.}~\bibnamefont
  {Su}}, \bibinfo {author} {\bibfnamefont {J.}~\bibnamefont {Xing}}, \bibinfo
  {author} {\bibfnamefont {Q.}~\bibnamefont {Xiong}}, \ and\ \bibinfo {author}
  {\bibfnamefont {H.~S.}\ \bibnamefont {Nguyen}},\ }\bibfield  {title}
  {\enquote {\bibinfo {title} {Unveiling the enhancement of spontaneous
  emission at exceptional points},}\ }\href {\doibase
  10.1103/PhysRevLett.129.083602} {\bibfield  {journal} {\bibinfo  {journal}
  {Phys. Rev. Lett.}\ }\textbf {\bibinfo {volume} {129}},\ \bibinfo {pages}
  {083602} (\bibinfo {year} {2022})}\BibitemShut {NoStop}%
\bibitem [{\citenamefont {Rotter}(2009)}]{PhaseRidity}%
  \BibitemOpen
  \bibfield  {author} {\bibinfo {author} {\bibfnamefont {I.}~\bibnamefont
  {Rotter}},\ }\bibfield  {title} {\enquote {\bibinfo {title} {A non-hermitian
  hamilton operator and the physics of open quantum systems},}\ }\href
  {https://iopscience.iop.org/article/10.1088/1751-8113/42/15/153001}
  {\bibfield  {journal} {\bibinfo  {journal} {J. Phys. A: Math. Theor.}\
  }\textbf {\bibinfo {volume} {42}},\ \bibinfo {pages} {153001} (\bibinfo
  {year} {2009})}\BibitemShut {NoStop}%
\bibitem [{\citenamefont {Soluyanov}\ and\ \citenamefont
  {Vanderbilt}(2012)}]{ParellelTransport}%
  \BibitemOpen
  \bibfield  {author} {\bibinfo {author} {\bibfnamefont {A.~A.}\ \bibnamefont
  {Soluyanov}}\ and\ \bibinfo {author} {\bibfnamefont {D.}~\bibnamefont
  {Vanderbilt}},\ }\bibfield  {title} {\enquote {\bibinfo {title} {Smooth gauge
  for topological insulators},}\ }\href {\doibase 10.1103/PhysRevB.85.115415}
  {\bibfield  {journal} {\bibinfo  {journal} {Phys. Rev. B}\ }\textbf {\bibinfo
  {volume} {85}},\ \bibinfo {pages} {115415} (\bibinfo {year}
  {2012})}\BibitemShut {NoStop}%
\bibitem [{\citenamefont {Agrawal}(1987)}]{nonlinear_gain1}%
  \BibitemOpen
  \bibfield  {author} {\bibinfo {author} {\bibfnamefont {G.}~\bibnamefont
  {Agrawal}},\ }\bibfield  {title} {\enquote {\bibinfo {title} {Gain
  nonlinearities in semiconductor lasers: Theory and application to distributed
  feedback lasers},}\ }\href {\doibase 10.1109/JQE.1987.1073406} {\bibfield
  {journal} {\bibinfo  {journal} {IEEE J. Quantum Electron.}\ }\textbf
  {\bibinfo {volume} {23}},\ \bibinfo {pages} {860} (\bibinfo {year}
  {1987})}\BibitemShut {NoStop}%
\bibitem [{\citenamefont {Duan}\ \emph {et~al.}(1992)\citenamefont {Duan},
  \citenamefont {Gallion},\ and\ \citenamefont {Agrawal}}]{nonlinear_gain2}%
  \BibitemOpen
  \bibfield  {author} {\bibinfo {author} {\bibfnamefont {G.~H.}\ \bibnamefont
  {Duan}}, \bibinfo {author} {\bibfnamefont {P.}~\bibnamefont {Gallion}}, \
  and\ \bibinfo {author} {\bibfnamefont {G.~P.}\ \bibnamefont {Agrawal}},\
  }\bibfield  {title} {\enquote {\bibinfo {title} {Effective nonlinear gain in
  semiconductor lasers},}\ }\href {\doibase 10.1109/68.122371} {\bibfield
  {journal} {\bibinfo  {journal} {IEEE Photon. Technol. Lett.}\ }\textbf
  {\bibinfo {volume} {4}},\ \bibinfo {pages} {218} (\bibinfo {year}
  {1992})}\BibitemShut {NoStop}%
\bibitem [{\citenamefont {Zdzislaw}(2005)}]{MCT}%
  \BibitemOpen
  \bibfield  {author} {\bibinfo {author} {\bibfnamefont {B.}~\bibnamefont
  {Zdzislaw}},\ }\href@noop {} {\emph {\bibinfo {title} {Modern Control
  Theory}}}\ (\bibinfo  {publisher} {Springer},\ \bibinfo {year}
  {2005})\BibitemShut {NoStop}%
\bibitem [{\citenamefont {Bai}\ \emph {et~al.}(2023)\citenamefont {Bai},
  \citenamefont {Li}, \citenamefont {Liu}, \citenamefont {Fang}, \citenamefont
  {Wan},\ and\ \citenamefont {Xiao}}]{NEP1}%
  \BibitemOpen
  \bibfield  {author} {\bibinfo {author} {\bibfnamefont {K.}~\bibnamefont
  {Bai}}, \bibinfo {author} {\bibfnamefont {J.-Z.}\ \bibnamefont {Li}},
  \bibinfo {author} {\bibfnamefont {T.-R.}\ \bibnamefont {Liu}}, \bibinfo
  {author} {\bibfnamefont {L.}~\bibnamefont {Fang}}, \bibinfo {author}
  {\bibfnamefont {D.}~\bibnamefont {Wan}}, \ and\ \bibinfo {author}
  {\bibfnamefont {M.}~\bibnamefont {Xiao}},\ }\bibfield  {title} {\enquote
  {\bibinfo {title} {Nonlinear exceptional points with a complete basis in
  dynamics},}\ }\href {\doibase 10.1103/PhysRevLett.130.266901} {\bibfield
  {journal} {\bibinfo  {journal} {Phys. Rev. Lett.}\ }\textbf {\bibinfo
  {volume} {130}},\ \bibinfo {pages} {266901} (\bibinfo {year}
  {2023})}\BibitemShut {NoStop}%
\bibitem [{\citenamefont {Bai}\ \emph {et~al.}(2024)\citenamefont {Bai},
  \citenamefont {Liu}, \citenamefont {Fang}, \citenamefont {Li}, \citenamefont
  {Lin}, \citenamefont {Wan},\ and\ \citenamefont {Xiao}}]{NEP2}%
  \BibitemOpen
  \bibfield  {author} {\bibinfo {author} {\bibfnamefont {K.}~\bibnamefont
  {Bai}}, \bibinfo {author} {\bibfnamefont {T.-R.}\ \bibnamefont {Liu}},
  \bibinfo {author} {\bibfnamefont {L.}~\bibnamefont {Fang}}, \bibinfo {author}
  {\bibfnamefont {J.-Z.}\ \bibnamefont {Li}}, \bibinfo {author} {\bibfnamefont
  {C.}~\bibnamefont {Lin}}, \bibinfo {author} {\bibfnamefont {D.}~\bibnamefont
  {Wan}}, \ and\ \bibinfo {author} {\bibfnamefont {M.}~\bibnamefont {Xiao}},\
  }\bibfield  {title} {\enquote {\bibinfo {title} {Observation of nonlinear
  exceptional points with a complete basis in dynamics},}\ }\href {\doibase
  10.1103/PhysRevLett.132.073802} {\bibfield  {journal} {\bibinfo  {journal}
  {Phys. Rev. Lett.}\ }\textbf {\bibinfo {volume} {132}},\ \bibinfo {pages}
  {073802} (\bibinfo {year} {2024})}\BibitemShut {NoStop}%
\bibitem [{\citenamefont {Liu}\ \emph {et~al.}(2020)\citenamefont {Liu},
  \citenamefont {Han},\ and\ \citenamefont {Liu}}]{NHTI1}%
  \BibitemOpen
  \bibfield  {author} {\bibinfo {author} {\bibfnamefont {J.~S.}\ \bibnamefont
  {Liu}}, \bibinfo {author} {\bibfnamefont {Y.~Z.}\ \bibnamefont {Han}}, \ and\
  \bibinfo {author} {\bibfnamefont {C.~S.}\ \bibnamefont {Liu}},\ }\bibfield
  {title} {\enquote {\bibinfo {title} {A new way to construct topological
  invariants of non-hermitian systems with the non-hermitian skin effect},}\
  }\href {\doibase 10.1088/1674-1056/ab5937} {\bibfield  {journal} {\bibinfo
  {journal} {Chin. Phys. B}\ }\textbf {\bibinfo {volume} {29}},\ \bibinfo
  {pages} {010302} (\bibinfo {year} {2020})}\BibitemShut {NoStop}%
\bibitem [{\citenamefont {Jiang}\ \emph {et~al.}(2018)\citenamefont {Jiang},
  \citenamefont {Yang},\ and\ \citenamefont {Chen}}]{NHTI2}%
  \BibitemOpen
  \bibfield  {author} {\bibinfo {author} {\bibfnamefont {H.}~\bibnamefont
  {Jiang}}, \bibinfo {author} {\bibfnamefont {C.}~\bibnamefont {Yang}}, \ and\
  \bibinfo {author} {\bibfnamefont {S.}~\bibnamefont {Chen}},\ }\bibfield
  {title} {\enquote {\bibinfo {title} {Topological invariants and phase
  diagrams for one-dimensional two-band non-hermitian systems without chiral
  symmetry},}\ }\href {\doibase 10.1103/PhysRevA.98.052116} {\bibfield
  {journal} {\bibinfo  {journal} {Phys. Rev. A}\ }\textbf {\bibinfo {volume}
  {98}},\ \bibinfo {pages} {052116} (\bibinfo {year} {2018})}\BibitemShut
  {NoStop}%
\bibitem [{\citenamefont {Zak}(1989)}]{ZakPhase1}%
  \BibitemOpen
  \bibfield  {author} {\bibinfo {author} {\bibfnamefont {J.}~\bibnamefont
  {Zak}},\ }\bibfield  {title} {\enquote {\bibinfo {title} {Berry's phase for
  energy bands in solids},}\ }\href {\doibase 10.1103/PhysRevLett.62.2747}
  {\bibfield  {journal} {\bibinfo  {journal} {Phys. Rev. Lett.}\ }\textbf
  {\bibinfo {volume} {62}},\ \bibinfo {pages} {2747} (\bibinfo {year}
  {1989})}\BibitemShut {NoStop}%
\bibitem [{\citenamefont {Xiao}\ \emph {et~al.}(2014)\citenamefont {Xiao},
  \citenamefont {Zhang},\ and\ \citenamefont {Chan}}]{ZakPhase2}%
  \BibitemOpen
  \bibfield  {author} {\bibinfo {author} {\bibfnamefont {M.}~\bibnamefont
  {Xiao}}, \bibinfo {author} {\bibfnamefont {Z.~Q.}\ \bibnamefont {Zhang}}, \
  and\ \bibinfo {author} {\bibfnamefont {C.~T.}\ \bibnamefont {Chan}},\
  }\bibfield  {title} {\enquote {\bibinfo {title} {Surface impedance and bulk
  band geometric phases in one-dimensional systems},}\ }\href {\doibase
  10.1103/PhysRevX.4.021017} {\bibfield  {journal} {\bibinfo  {journal} {Phys.
  Rev. X}\ }\textbf {\bibinfo {volume} {4}},\ \bibinfo {pages} {021017}
  (\bibinfo {year} {2014})}\BibitemShut {NoStop}%
\bibitem [{\citenamefont {Yokomizo}\ and\ \citenamefont
  {Murakami}(2019)}]{1DNHchain1}%
  \BibitemOpen
  \bibfield  {author} {\bibinfo {author} {\bibfnamefont {K.}~\bibnamefont
  {Yokomizo}}\ and\ \bibinfo {author} {\bibfnamefont {S.}~\bibnamefont
  {Murakami}},\ }\bibfield  {title} {\enquote {\bibinfo {title} {Non-bloch band
  theory of non-hermitian systems},}\ }\href {\doibase
  10.1103/PhysRevLett.123.066404} {\bibfield  {journal} {\bibinfo  {journal}
  {Phys. Rev. Lett.}\ }\textbf {\bibinfo {volume} {123}},\ \bibinfo {pages}
  {066404} (\bibinfo {year} {2019})}\BibitemShut {NoStop}%
\bibitem [{\citenamefont {Yao}\ and\ \citenamefont {Wang}(2018)}]{1DNHchain2}%
  \BibitemOpen
  \bibfield  {author} {\bibinfo {author} {\bibfnamefont {S.}~\bibnamefont
  {Yao}}\ and\ \bibinfo {author} {\bibfnamefont {Z.}~\bibnamefont {Wang}},\
  }\bibfield  {title} {\enquote {\bibinfo {title} {Edge states and topological
  invariants of non-hermitian systems},}\ }\href {\doibase
  10.1103/PhysRevLett.121.086803} {\bibfield  {journal} {\bibinfo  {journal}
  {Phys. Rev. Lett.}\ }\textbf {\bibinfo {volume} {121}},\ \bibinfo {pages}
  {086803} (\bibinfo {year} {2018})}\BibitemShut {NoStop}%
\bibitem [{\citenamefont {Su}\ \emph {et~al.}(1979)\citenamefont {Su},
  \citenamefont {Schrieffer},\ and\ \citenamefont {Heeger}}]{SSHmodel}%
  \BibitemOpen
  \bibfield  {author} {\bibinfo {author} {\bibfnamefont {W.~P.}\ \bibnamefont
  {Su}}, \bibinfo {author} {\bibfnamefont {J.~R.}\ \bibnamefont {Schrieffer}},
  \ and\ \bibinfo {author} {\bibfnamefont {A.~J.}\ \bibnamefont {Heeger}},\
  }\bibfield  {title} {\enquote {\bibinfo {title} {Solitons in
  polyacetylene},}\ }\href {\doibase 10.1103/PhysRevLett.42.1698} {\bibfield
  {journal} {\bibinfo  {journal} {Phys. Rev. Lett.}\ }\textbf {\bibinfo
  {volume} {42}},\ \bibinfo {pages} {1698} (\bibinfo {year}
  {1979})}\BibitemShut {NoStop}%
\bibitem [{\citenamefont {Manzano}(2020)}]{manzano2020aip}%
  \BibitemOpen
  \bibfield  {author} {\bibinfo {author} {\bibfnamefont {D.}~\bibnamefont
  {Manzano}},\ }\bibfield  {title} {\enquote {\bibinfo {title} {{A short
  introduction to the Lindblad master equation}},}\ }\href {\doibase
  10.1063/1.5115323} {\bibfield  {journal} {\bibinfo  {journal} {AIP Adv.}\
  }\textbf {\bibinfo {volume} {10}},\ \bibinfo {pages} {025106} (\bibinfo
  {year} {2020})}\BibitemShut {NoStop}%
\bibitem [{\citenamefont {Mittal}\ \emph {et~al.}(2019)\citenamefont {Mittal},
  \citenamefont {Orre}, \citenamefont {Zhu}, \citenamefont {Gorlach},
  \citenamefont {Poddubny},\ and\ \citenamefont {Hafezi}}]{pQTP}%
  \BibitemOpen
  \bibfield  {author} {\bibinfo {author} {\bibfnamefont {S.}~\bibnamefont
  {Mittal}}, \bibinfo {author} {\bibfnamefont {V.~V.}\ \bibnamefont {Orre}},
  \bibinfo {author} {\bibfnamefont {G.~Y.}\ \bibnamefont {Zhu}}, \bibinfo
  {author} {\bibfnamefont {M.~A.}\ \bibnamefont {Gorlach}}, \bibinfo {author}
  {\bibfnamefont {A.}~\bibnamefont {Poddubny}}, \ and\ \bibinfo {author}
  {\bibfnamefont {M.}~\bibnamefont {Hafezi}},\ }\bibfield  {title} {\enquote
  {\bibinfo {title} {Photonic quadrupole topological phases},}\ }\href
  {https://www.nature.com/articles/s41566-019-0452-0} {\bibfield  {journal}
  {\bibinfo  {journal} {Nat. Photon.}\ }\textbf {\bibinfo {volume} {13}},\
  \bibinfo {pages} {692} (\bibinfo {year} {2019})}\BibitemShut {NoStop}%
\bibitem [{\citenamefont {Hu}\ \emph {et~al.}(2021)\citenamefont {Hu},
  \citenamefont {Zhang}, \citenamefont {Zhang}, \citenamefont {Zheng},
  \citenamefont {Xiong}, \citenamefont {Yue}, \citenamefont {Wang},
  \citenamefont {Xu}, \citenamefont {Cheng}, \citenamefont {Liu},\ and\
  \citenamefont {Christensen}}]{NHWisper}%
  \BibitemOpen
  \bibfield  {author} {\bibinfo {author} {\bibfnamefont {B.}~\bibnamefont
  {Hu}}, \bibinfo {author} {\bibfnamefont {Z.}~\bibnamefont {Zhang}}, \bibinfo
  {author} {\bibfnamefont {H.}~\bibnamefont {Zhang}}, \bibinfo {author}
  {\bibfnamefont {L.}~\bibnamefont {Zheng}}, \bibinfo {author} {\bibfnamefont
  {W.}~\bibnamefont {Xiong}}, \bibinfo {author} {\bibfnamefont
  {Z.}~\bibnamefont {Yue}}, \bibinfo {author} {\bibfnamefont {X.}~\bibnamefont
  {Wang}}, \bibinfo {author} {\bibfnamefont {J.}~\bibnamefont {Xu}}, \bibinfo
  {author} {\bibfnamefont {Y.}~\bibnamefont {Cheng}}, \bibinfo {author}
  {\bibfnamefont {X.}~\bibnamefont {Liu}}, \ and\ \bibinfo {author}
  {\bibfnamefont {J.}~\bibnamefont {Christensen}},\ }\bibfield  {title}
  {\enquote {\bibinfo {title} {Non-hermitian topological whispering gallery},}\
  }\href {https://www.nature.com/articles/s41586-021-03833-4} {\bibfield
  {journal} {\bibinfo  {journal} {Nature}\ }\textbf {\bibinfo {volume} {597}},\
  \bibinfo {pages} {655} (\bibinfo {year} {2021})}\BibitemShut {NoStop}%
\bibitem [{\citenamefont {Xu}\ and\ \citenamefont {Pu}(2019)}]{hpu2019prl}%
  \BibitemOpen
  \bibfield  {author} {\bibinfo {author} {\bibfnamefont {Y.}~\bibnamefont
  {Xu}}\ and\ \bibinfo {author} {\bibfnamefont {H.}~\bibnamefont {Pu}},\
  }\bibfield  {title} {\enquote {\bibinfo {title} {Emergent universality in a
  quantum tricritical dicke model},}\ }\href {\doibase
  10.1103/PhysRevLett.122.193201} {\bibfield  {journal} {\bibinfo  {journal}
  {Phys. Rev. Lett.}\ }\textbf {\bibinfo {volume} {122}},\ \bibinfo {pages}
  {193201} (\bibinfo {year} {2019})}\BibitemShut {NoStop}%
\bibitem [{\citenamefont {Dimer}\ \emph {et~al.}(2007)\citenamefont {Dimer},
  \citenamefont {Estienne}, \citenamefont {Parkins},\ and\ \citenamefont
  {Carmichael}}]{hjc2007pra}%
  \BibitemOpen
  \bibfield  {author} {\bibinfo {author} {\bibfnamefont {F.}~\bibnamefont
  {Dimer}}, \bibinfo {author} {\bibfnamefont {B.}~\bibnamefont {Estienne}},
  \bibinfo {author} {\bibfnamefont {A.~S.}\ \bibnamefont {Parkins}}, \ and\
  \bibinfo {author} {\bibfnamefont {H.~J.}\ \bibnamefont {Carmichael}},\
  }\bibfield  {title} {\enquote {\bibinfo {title} {Proposed realization of the
  dicke-model quantum phase transition in an optical cavity qed system},}\
  }\href {\doibase 10.1103/PhysRevA.75.013804} {\bibfield  {journal} {\bibinfo
  {journal} {Phys. Rev. A}\ }\textbf {\bibinfo {volume} {75}},\ \bibinfo
  {pages} {013804} (\bibinfo {year} {2007})}\BibitemShut {NoStop}%
\bibitem [{\citenamefont {Exner}(2011)}]{pathintegral}%
  \BibitemOpen
  \bibfield  {author} {\bibinfo {author} {\bibfnamefont {P.}~\bibnamefont
  {Exner}},\ }\href@noop {} {\emph {\bibinfo {title} {Open quantum systems and
  Feynman integrals}}}\ (\bibinfo  {publisher} {Springer, New York},\ \bibinfo
  {year} {2011})\BibitemShut {NoStop}%
\bibitem [{\citenamefont {Jauho}\ and\ \citenamefont {Haug}(2008)}]{koch}%
  \BibitemOpen
  \bibfield  {author} {\bibinfo {author} {\bibfnamefont {A.-P.}\ \bibnamefont
  {Jauho}}\ and\ \bibinfo {author} {\bibfnamefont {H.}~\bibnamefont {Haug}},\
  }\href@noop {} {\emph {\bibinfo {title} {Quantum Kinetics in Transport and
  Optics of Semiconductors}}}\ (\bibinfo  {publisher} {Springer},\ \bibinfo
  {year} {2008})\BibitemShut {NoStop}%
\bibitem [{\citenamefont {Wu}\ \emph {et~al.}(2010)\citenamefont {Wu},
  \citenamefont {Jiang},\ and\ \citenamefont {Weng}}]{phrep}%
  \BibitemOpen
  \bibfield  {author} {\bibinfo {author} {\bibfnamefont {M.~W.}\ \bibnamefont
  {Wu}}, \bibinfo {author} {\bibfnamefont {J.-H.}\ \bibnamefont {Jiang}}, \
  and\ \bibinfo {author} {\bibfnamefont {M.~Q.}\ \bibnamefont {Weng}},\
  }\bibfield  {title} {\enquote {\bibinfo {title} {Spin dynamics in
  semiconductors},}\ }\href
  {https://www.sciencedirect.com/science/article/abs/pii/S0370157310000955}
  {\bibfield  {journal} {\bibinfo  {journal} {Phys. Rep.}\ }\textbf {\bibinfo
  {volume} {493}},\ \bibinfo {pages} {61} (\bibinfo {year} {2010})}\BibitemShut
  {NoStop}%
\end{thebibliography}%

\end{document}